\documentclass[apj,iop,twocolumn]{emulateapj_notes}
\usepackage{tikz}
\usepackage{pict2e}
\usepackage{lipsum}
\usepackage[intlimits]{amsmath}
\usepackage{amsmath}
\usepackage{mathptmx}
\usepackage{natbib}
\usepackage{bm}
\usepackage[scaled=.90]{helvet}
\usepackage{courier}
\usepackage{epsfig}
\usepackage{graphicx}
\usepackage{natbib,twoopt}
\usepackage[breaklinks=true]{hyperref} 
\bibpunct{(}{)}{;}{a}{}{,}             
\makeatletter
  \newcommandtwoopt{\citeads}[3][][]{\href{http://adsabs.harvard.edu/abs/#3}%
    {\def\hyper@linkstart##1##2{}%
     \let\hyper@linkend\@empty\citealp[#1][#2]{#3}}}
  \newcommandtwoopt{\citepads}[3][][]{\href{http://adsabs.harvard.edu/abs/#3}%
    {\def\hyper@linkstart##1##2{}%
     \let\hyper@linkend\@empty\citep[#1][#2]{#3}}}
  \newcommandtwoopt{\citetads}[3][][]{\href{http://adsabs.harvard.edu/abs/#3}%
    {\def\hyper@linkstart##1##2{}%
     \let\hyper@linkend\@empty\citet[#1][#2]{#3}}}
  \newcommandtwoopt{\citeyearads}[3][][]%
    {\href{http://adsabs.harvard.edu/abs/#3}
    {\def\hyper@linkstart##1##2{}%
     \let\hyper@linkend\@empty\citeyear[#1][#2]{#3}}}
\makeatother

\def\dtb{\delta{T}_b^\nu}
\def\nm{2\ell+1}
\def\nmb{\bigl(2\ell+1\bigr)}
\def\pfppp{P_{f_\parallel}}

\def\bfig{\begin{figure}}
\def\efig{\end{figure}}
\def\ints{\int\!}
\def\alm{a_{\lm}}

\def\gsim{\mathrel{\rlap{\lower2pt\hbox{\hskip0pt\small$\sim$}}
\raise2pt\hbox{\small $>$}}}
\def\lsim{\mathrel{\rlap{\lower2pt\hbox{\hskip0pt\small$\sim$}}
\raise2pt\hbox{\small $<$}}}     \def\bq{\begin{equation}}

\def\jl{j_\ell}
\def\cl{C_\ell}
\def\dl{\mathcal{D}_\ell}
\def\dlt{\mathcal{D}_{3000}}
\def\l{\ell}
\def\lm{\ell{m}}

\def\vel{{\bf v}}
\def\velm{\rm v}

\def\r{{\bf r}}
\def\q{{\bf q}}
\def\qk{{\bf{q}}}
\def\nhat{\hat{\bm \gamma}}
\def\khat{\hat{\k}}

\def\k{{\bm{k}}}
\def\d3k{\frac{\!d^3\k\!}{\!(2\pi)^3\!}}
\def\qprp{q_{\perp}}
\def\qpar{q_{\parallel}}
\def\qprpv{\q_{\perp}}

\def\dt{{\Delta}T(\nhat)}
\def\dtprp{{\Delta}T_\perp(\nhat)}
\def\dtpar{{\Delta}T_\parallel(\nhat)}
\def\nu{v}
\def\sn{\bigl({\rm S}/{\rm N}\bigr)^2}

\newcommand{\beq}{\begin{equation}}
\newcommand{\eeq}{\end{equation}}
\newcommand{\beqa}{\begin{eqnarray}}
\newcommand{\eeqa}{\end{eqnarray}}

\newcommand{\taua}{\langle\tau\rangle}

\newcommand{\gh}{{\hat{\gamma}}}

\newcommand{\vgh}{{\hat{\boldsymbol\gamma}}}

\newcommand{\beal}{\begin{align}}

\newcommand{\enal}{\end{align}}

\newcommand{\bspl}{\begin{split}}

\newcommand{\espl}{\end{split}}

\newcommand{\bsub}{\begin{subequations}}

\newcommand{\esub}{\end{subequations}}

\newcommand{\bmulti}{\begin{multline}}   %

\newcommand{\beqm}{\begin{mathletters}}   %

\newcommand{\eeqm}{\end{mathletters}}   %

\newcommand{\sigT}{\sigma_{\rm T}}

\slugcomment{Submitted to ApJ}

\shorttitle{Reionization kSZ at $10\lsim\ell\lsim{3000}$}
\shortauthors{{Alvarez}}

\begin{document}

\graphicspath{{./figures/}}
\def\figurewidth{0.47\textwidth}
\def\figurewidthn{0.45\textwidth}
\def\FigureOne{
\begin{figure}[t]
\begin{center}
\includegraphics[trim = 0 20.5 0 13, width=0.49\textwidth]{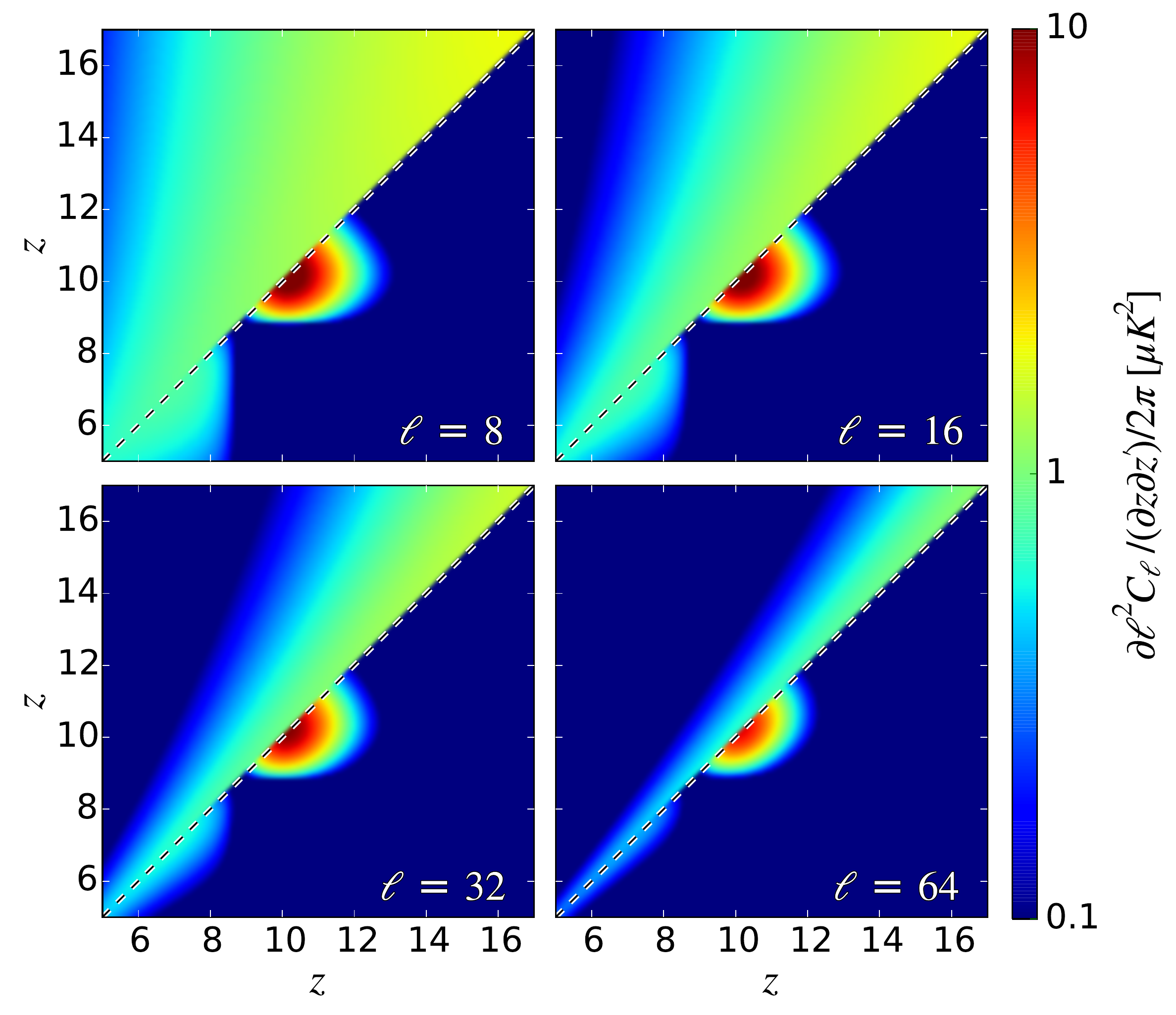}
\end{center}
\caption{Differential contribution to the angular power spectrum of the Doppler effect (equation \ref{eq:dcl}) at selected $\ell$ values, as labelled. The top-left half of each panel shows the contribution to the total angular power spectrum, assuming that the ionized fraction does not change, $x=1$, while the bottom-right shows the case for reionization, with a redshift dependence given by equation (\ref{eq:tanh}), with $\Delta{z}=0.5$ and $z_r=10$.}
\label{fig:dcldzdz}
\end{figure}
}

\def\FigureTwo{
\begin{figure}[t]
\begin{center}
\includegraphics[trim = 0 30 0 0, width=0.44\textwidth]{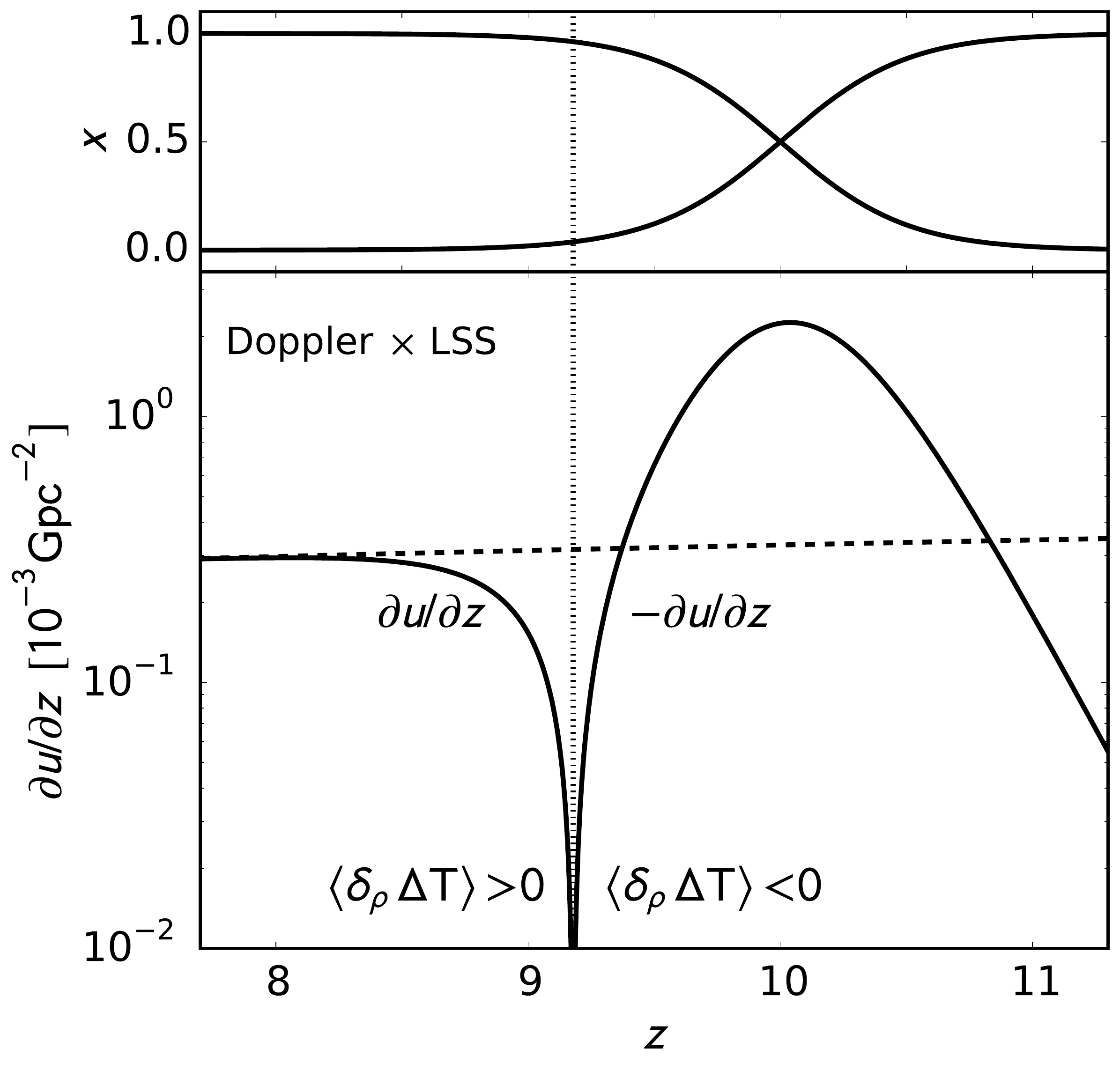}
\end{center}
\caption{Effect of reionization on the sign of density--Doppler correlations for the same reionization history (shown in the top panel) as in Figure \ref{fig:dcldzdz}. The dependence of $\partial{u}/\partial{z}$ on redshift is shown in the bottom panel. To the left of the vertical dotted line, $\partial{u}/\partial{z}>0$ and overdense regions will appear as positive kSZ temperature fluctuations, $\langle\delta_\rho\Delta{\rm T}\rangle>0$. At higher redshift, the sign of $\partial{u}/\partial{z}$ changes, and regions falling toward an overdense region will be nearer to the observer when the universe is more highly ionized, and the net effect will be a redshift of the CMB photons, so that $\langle\delta_\rho\Delta{\rm T}\rangle<0$. The dashed line shows $\partial{u}/\partial{z}$ for a constant reionization history, $x=1$.}
\label{fig:dudz}
\end{figure}
}

\def\FigureThree{
\begin{figure}[t]
\begin{center}
\includegraphics[trim = 0 30 0 10, width=\figurewidth]{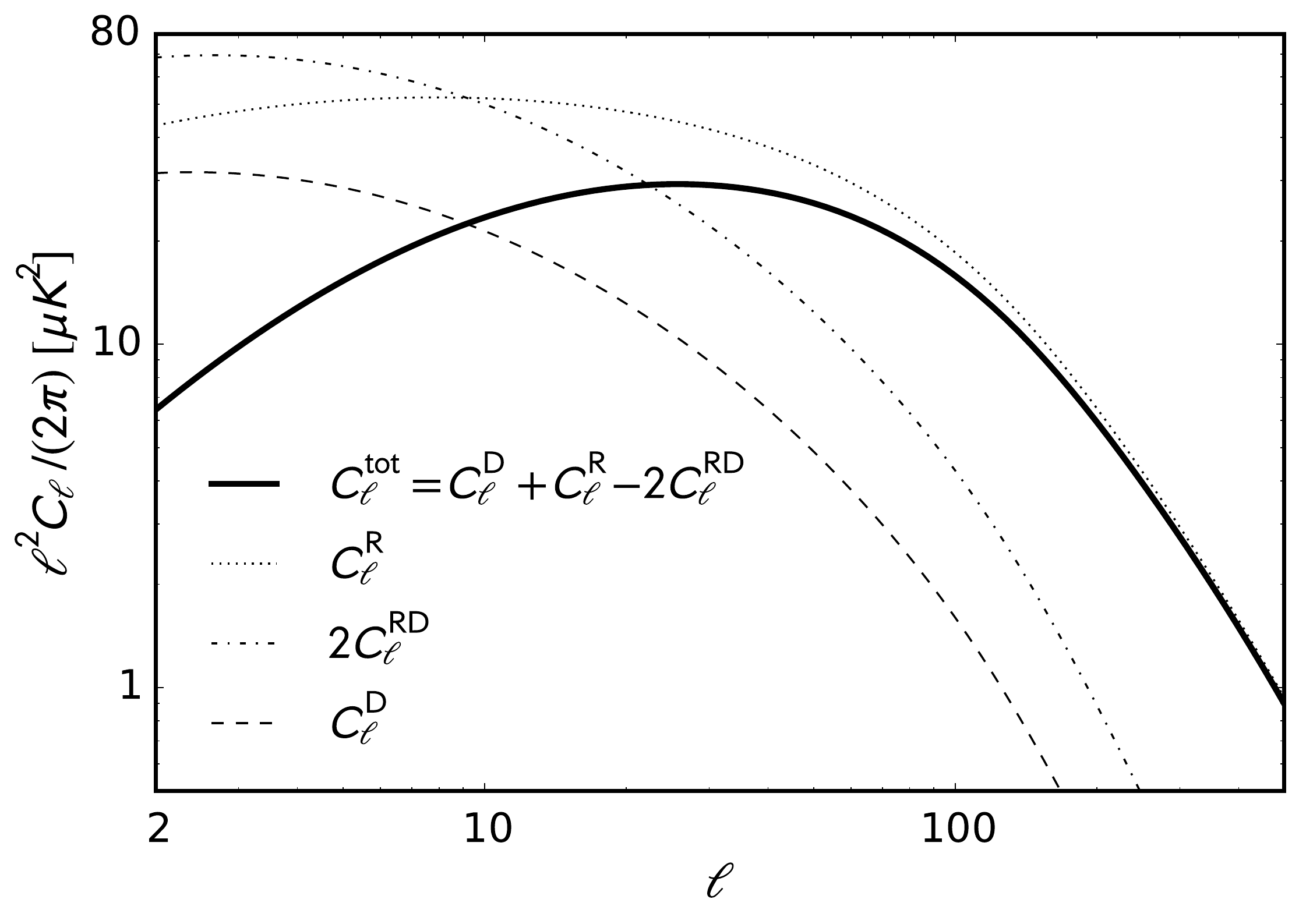}
\end{center}
\caption{Angular power spectrum of Doppler effect for instantaneous reionization occurring at $z=10$. The dotted curve labeled $\cl^R$ corresponds to the contribution from velocity fluctuations projected onto the re-scattering surface. Note that those fluctuations are larger and peak at smaller scales than the integrated term, $\cl^D$. The factor $2\cl^{\rm RD}$ accounts for the partial cancellation of the re-scattering surface fluctuations by flows on the near side. }
\label{fig:instantaneous}
\end{figure}
}

\def\FigureFour{
\begin{figure}
\begin{center}
\includegraphics[trim = 0 30 0 0, width=\figurewidth]{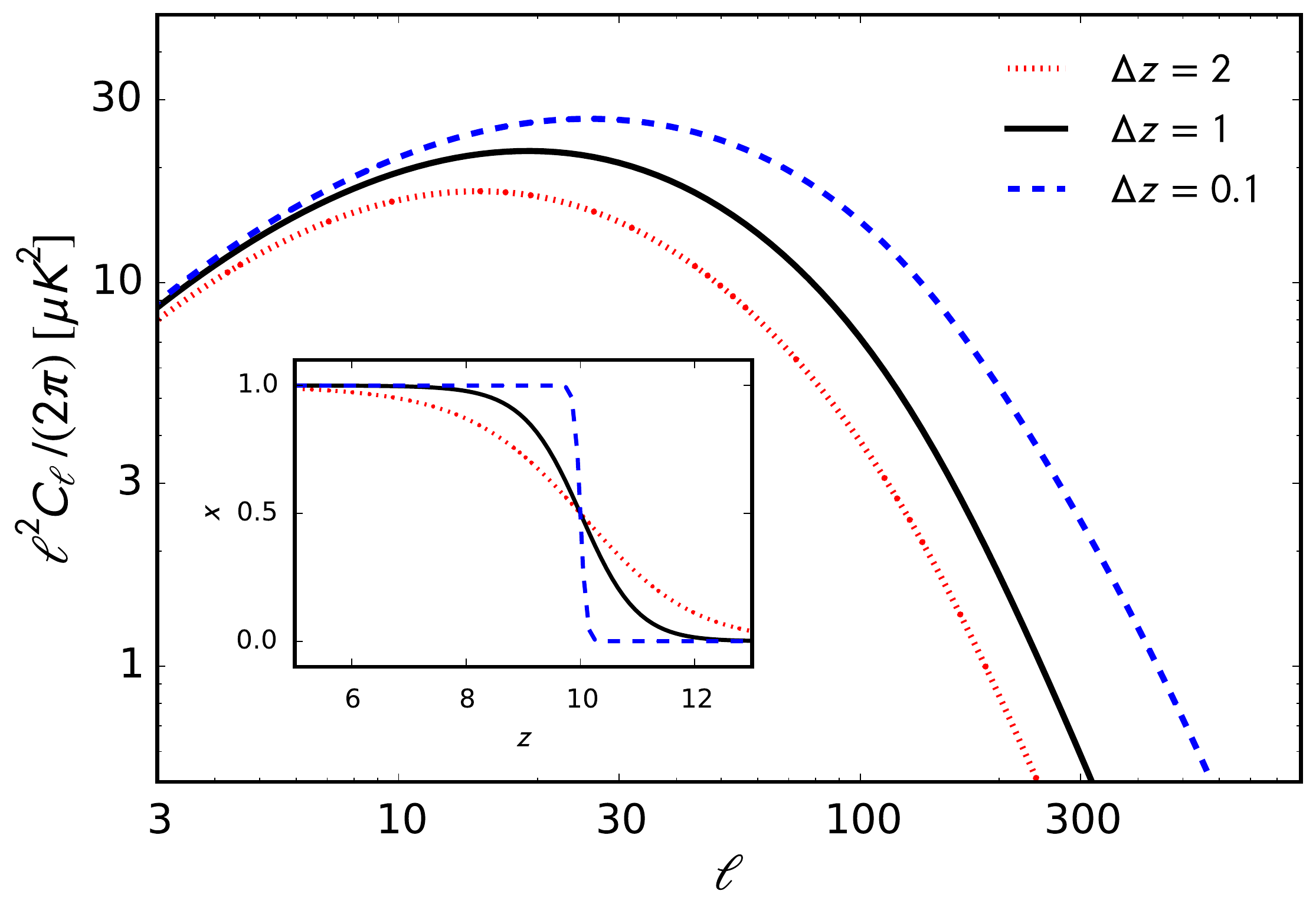}
\end{center}
\caption{Angular power spectrum of Doppler effect for different reionization histories given by equation (\ref{eq:tanh}), with $z_r=10$ and $\Delta\!z_{\rm reion}=0.1, 1$, and 2, as labelled. The inset shows the reionization history for each case. The shorter the duration of reionization at fixed redshift, the larger the amplitude of the power spectrum.}
\label{fig:cls_anl}
\end{figure}
}

\def\FigureFive{
\begin{figure}
\begin{center}
\includegraphics[trim = 0 30 0 0, width=\figurewidth]{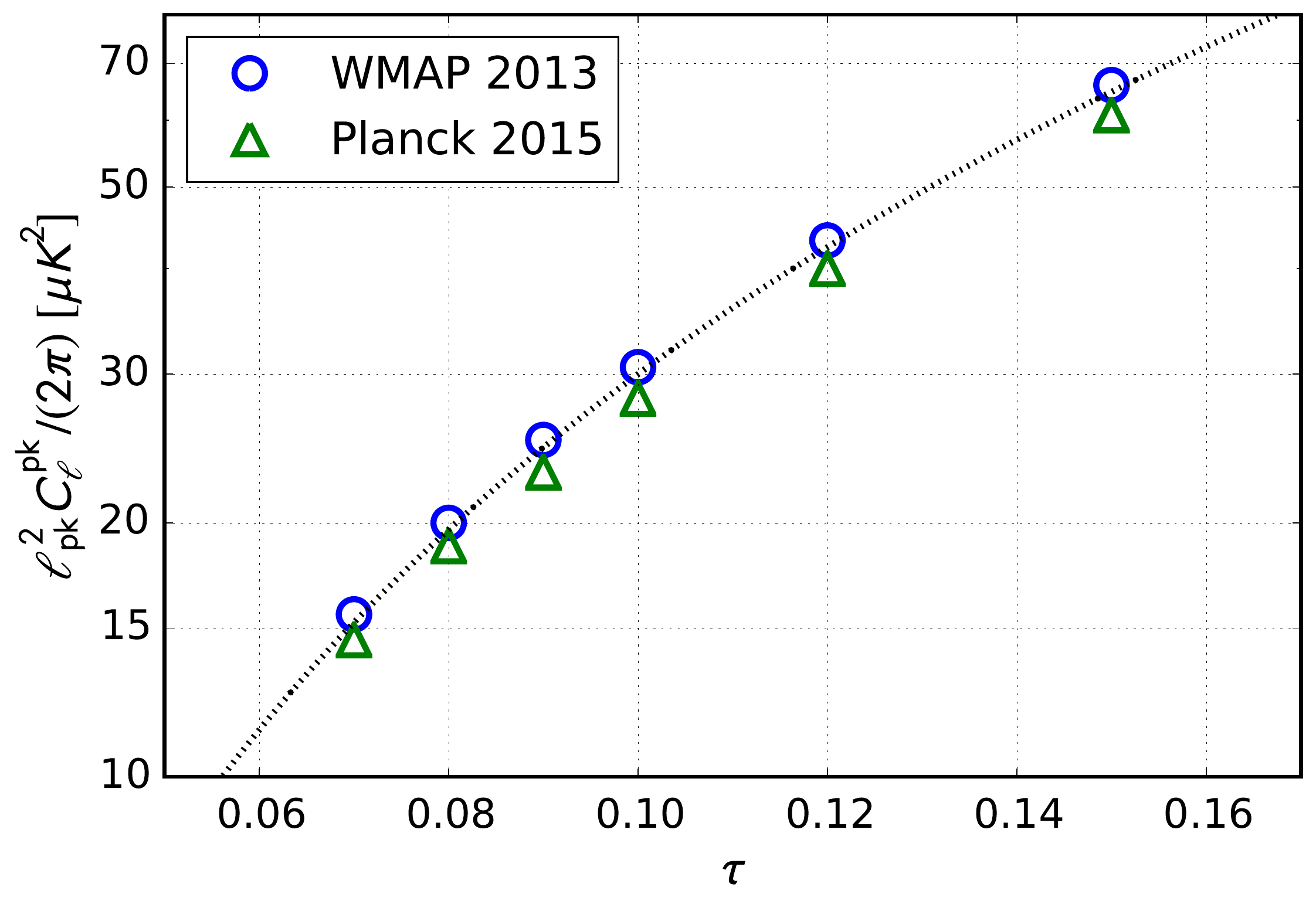}
\end{center}
\caption{Peak amplitude of Doppler effect power spectrum vs. $\tau$ for the two fiducial cosmological parameter sets considered. The dotted line corresponds to the fit given in equation (\ref{fig:fit}).}
\label{fig:clpk}
\end{figure}
}

\def\FigureSix{
\begin{figure}[t]
\begin{center}
\includegraphics[width=0.5\textwidth]{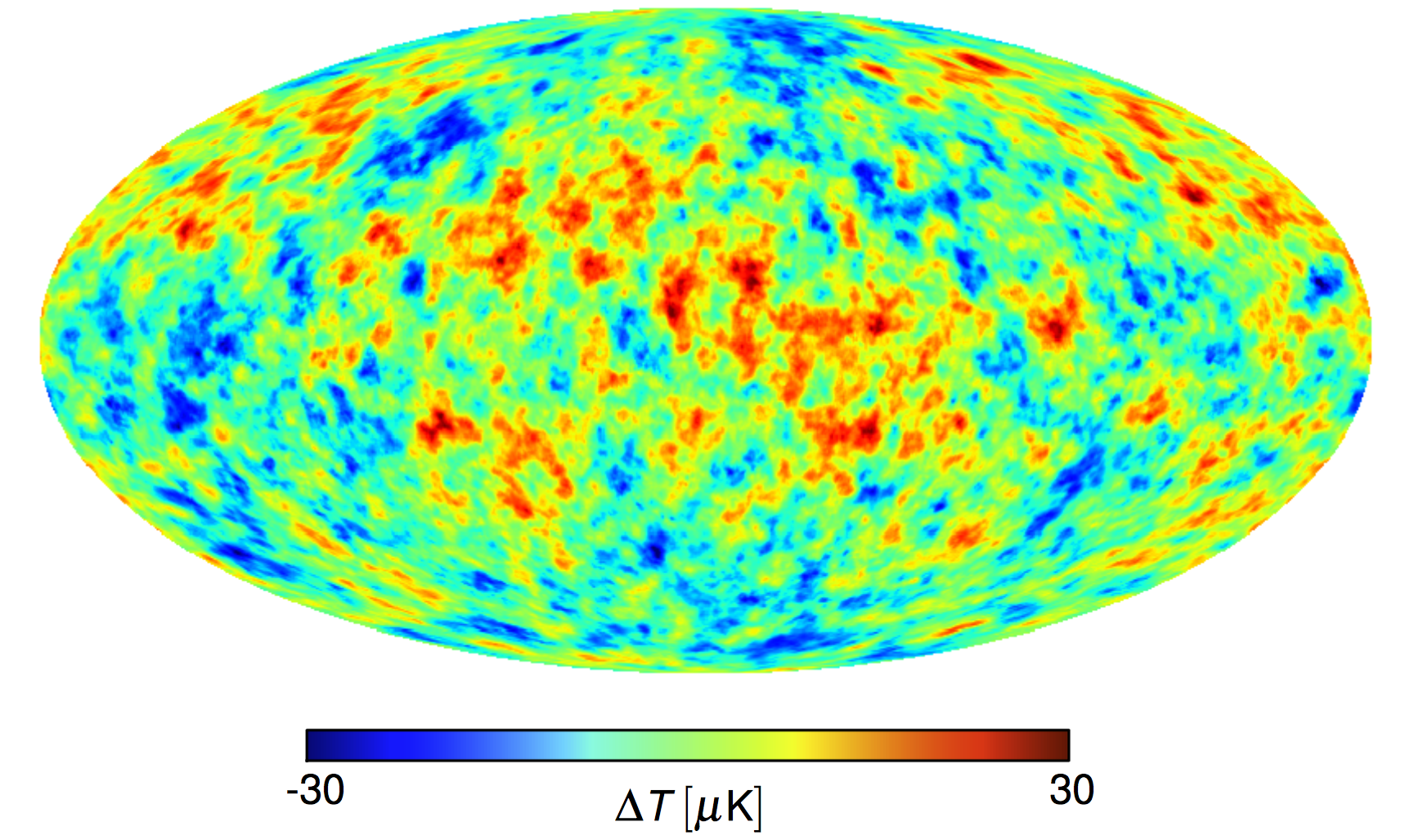}
\vspace{-0.4cm}
\end{center}
\caption{A full sky map produced with the map making procedure described in \S\ref{subsec:simmaps} for a patchy reionization simulation with $4096^3$ resolution elements in a periodic volume 8~Gpc$/h$ across with $\tau=0.09$.}
\vspace{0.2cm}
\label{fig:asm}
\end{figure}
}

\def\FigureSeven{
\begin{figure}
\begin{center}
\includegraphics[width=0.47\textwidth, trim= 0 0 0 0, clip=true]{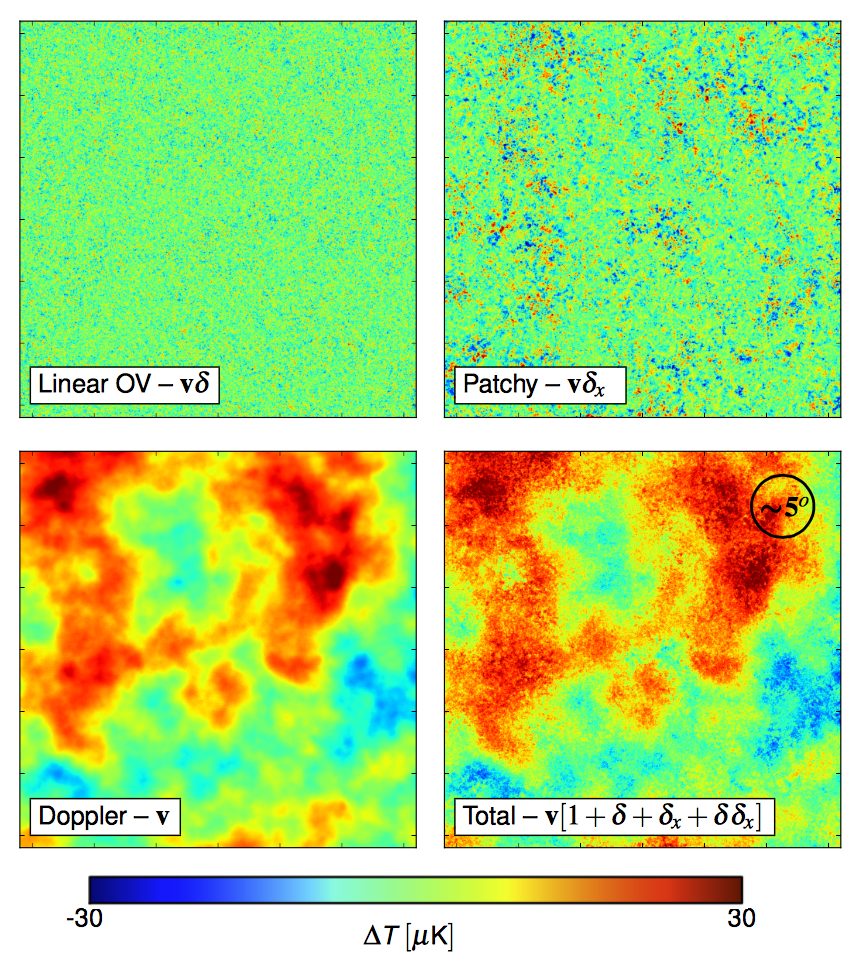}
%
%
%
%
%
%
%
%
\vspace{-0.4cm}
\end{center}
\caption{kSZ temperature fluctuations in a $32\times\!32$ degree region of the full sky map shown in Figure \ref{fig:asm}. The total is shown in the lower right panel, while the others correspond to the individual terms in the integrand  of equation (\ref{eq:ksz2}). The absolute value of the large scale velocity fluctuations (lower left)  correlate with regions of increased small scale power, most easily seen by comparing hot and cold regions in the lower left (Doppler) panel to the same regions in the top right (patchy) panel.}
\label{fig:maps}
\end{figure}
}

\def\FigureEight{
\begin{figure}
\begin{center}
\includegraphics[trim = 0 30 0 0, width=\figurewidth]{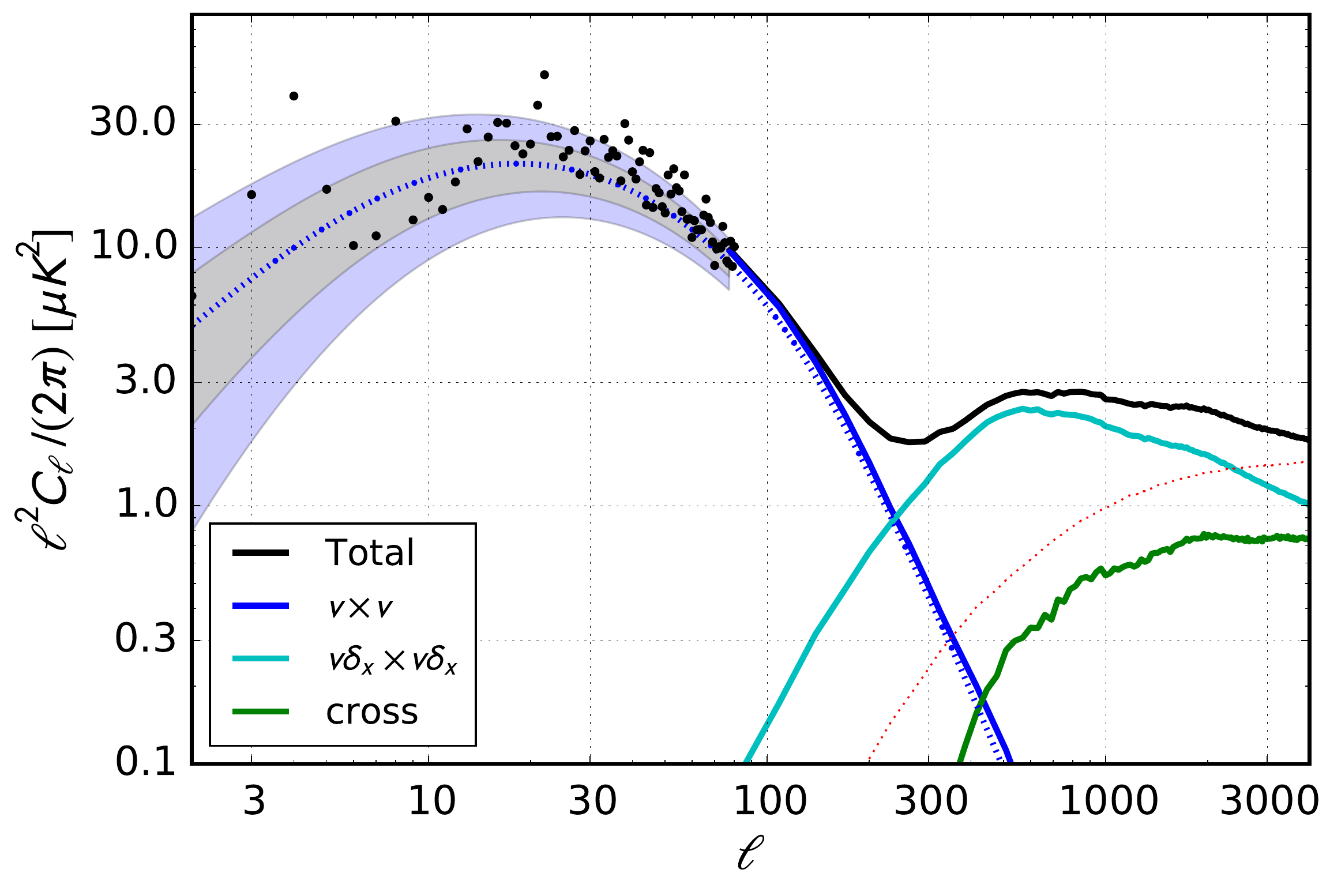}
\end{center}
\caption{Angular power spectra calculated from the maps shown in Figure \ref{fig:maps}, as labeled. The smooth red curve peaking at $\ell\sim 20-30$ is the Doppler component obtained with equation (\ref{eq:cl}) and the approximation of equation (\ref{eq:f1}), where $u(z)$ is determined from the mean ionization history in the simulation. Shaded regions show 1 and 2--$\sigma$ cosmic variance limits, assuming a $\chi^2$ distribution with mean $\cl$ and $2\ell+1$ degrees of freedom.}
\label{fig:cls_sim}
\end{figure}
}

\def\FigureNine{
\begin{figure}
\begin{center}
\includegraphics[width=\figurewidth, clip=true, trim=0 0 0 0]{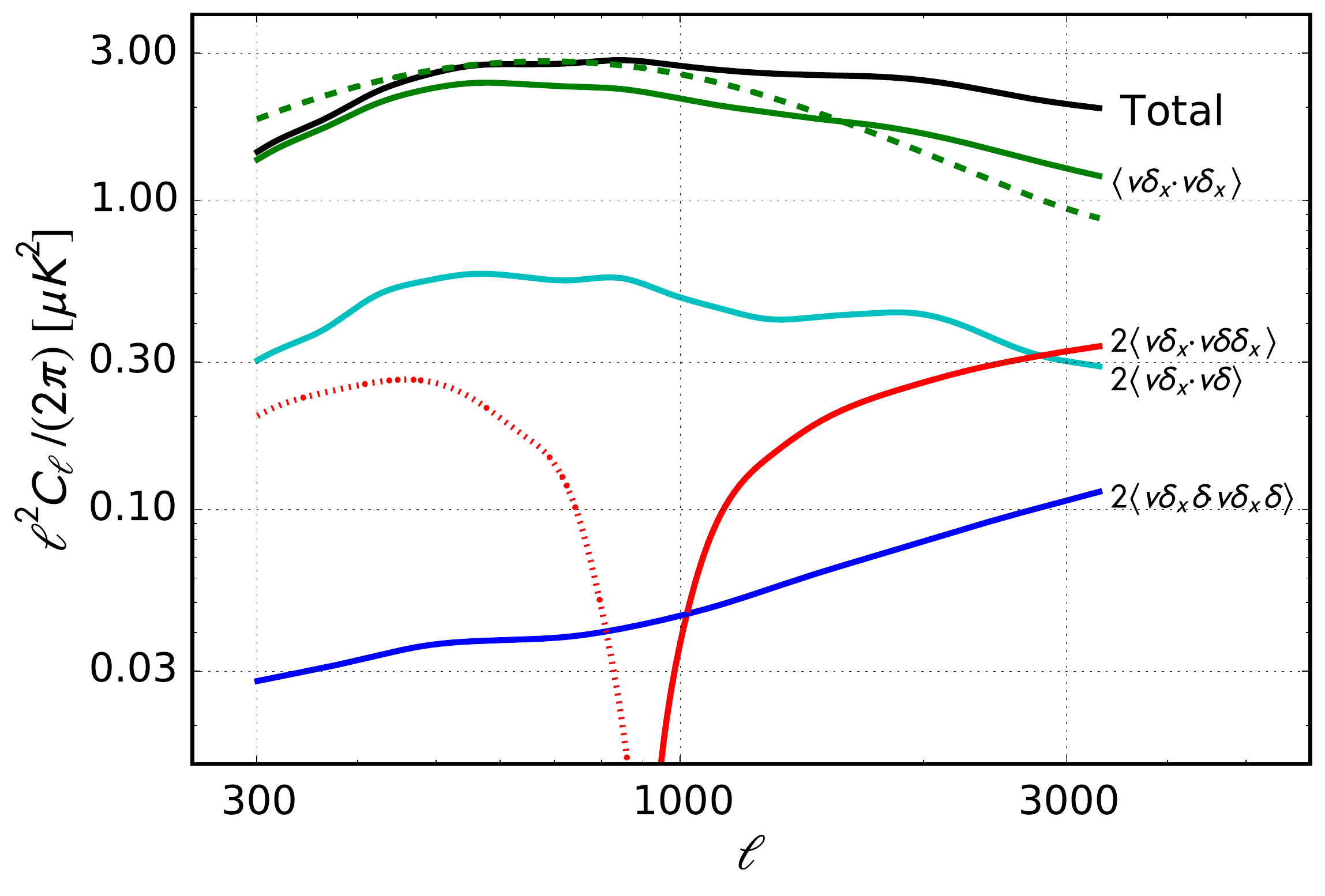}
\vspace{-0.3cm}
\end{center}
\caption{{\em left}: Deconstruction of patchy components of the kSZ map.  Total in this figure corresponds to the angular power spectrum of the map with `non--patchy' terms removed as in equation (\ref{eq:patchy}).}
\label{fig:terms}
\end{figure}
}

\def\FigureTen{
\begin{figure}
\begin{center}
\includegraphics[width=0.45\textwidth]{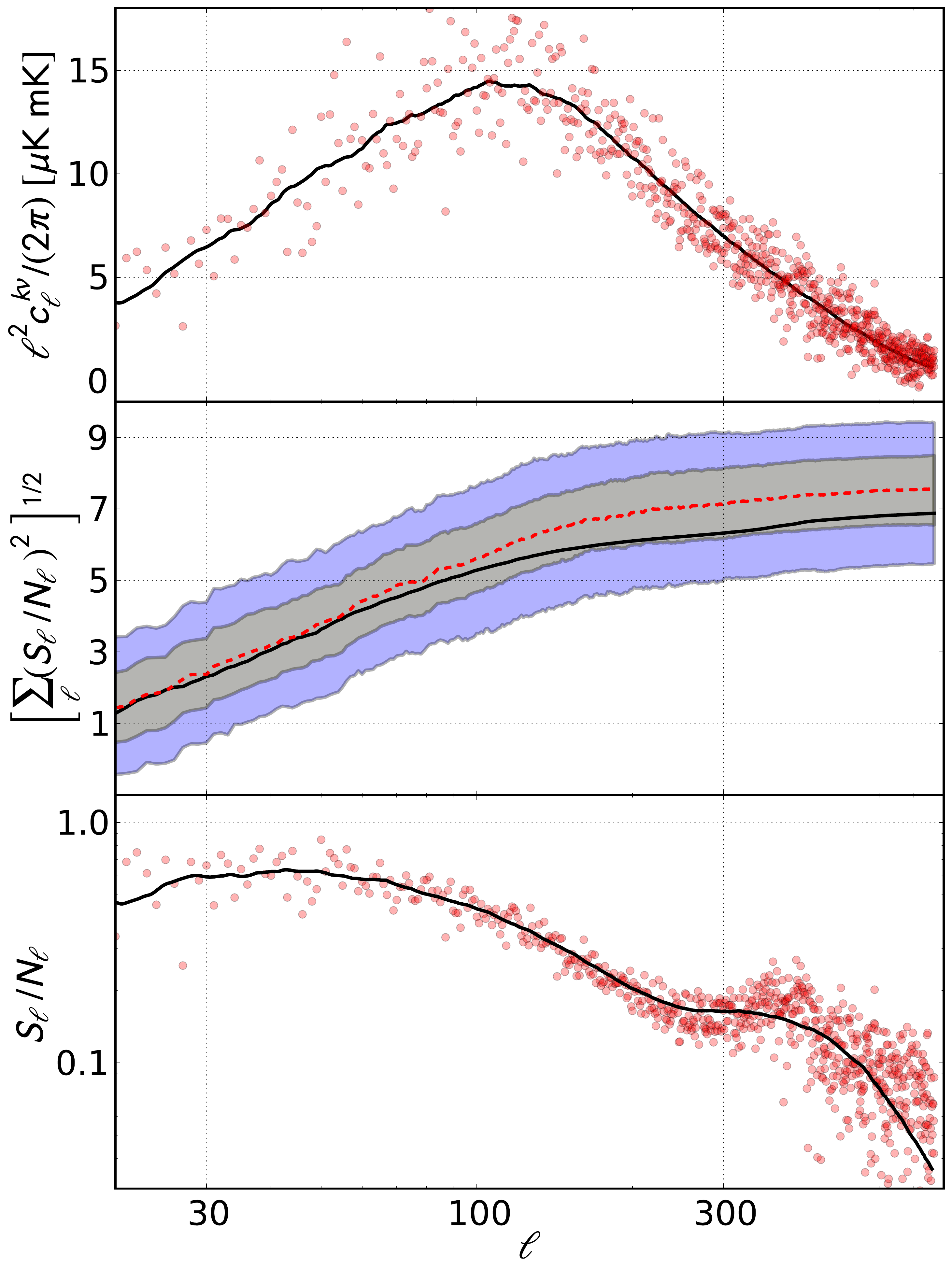}
\vspace{-0.3cm}
\end{center}
\caption{Angular power spectrum and signal--to--noise estimates for the CMB--21cm cross-correlation. {\em top}: Cross-correlation power spectrum of the kSZ and 21--cm map. Points show the individual multipole values while the solid line is a binned representation. The signal exhibits a strong positive peak corresponding to the peak of the matter power spectrum at ${0.01}{\rm Mpc}^{-1}$ at a distance of approximately $10\ {\rm Gpc}$, $\ell\!\sim 0.01\ {\rm Mpc}^{-1}\ 10^4\ {\rm Gpc} = 100$, as expected in typical reionization scenarios where highly--biased rare sources drive the growth of \ion{H}{2} regions. {\em middle}: Integrated signal to noise ratio estimate, using equation (\ref{eq:snrknu}; black line) compared to the results from the Monte Carlo calculations explained in \S\label{sec:21vz}. The red dashed line indicates the medium value for the power summed to that multipole, divided by the variance, while the shaded regions show the one and two--$\sigma$ ranges for the Monte Carlo results. {\em bottom}: signal--to--noise of individual $\ell$--modes in the kSZ--21cm cross-correlation, estimated using equation (\ref{eq:rknu}).}
\label{fig:cl21ksz}
\end{figure}
}

\def\FigureEleven{
\begin{figure}
\begin{center}
\includegraphics[width=\figurewidthn, trim = 0 0 0 0, clip=true]{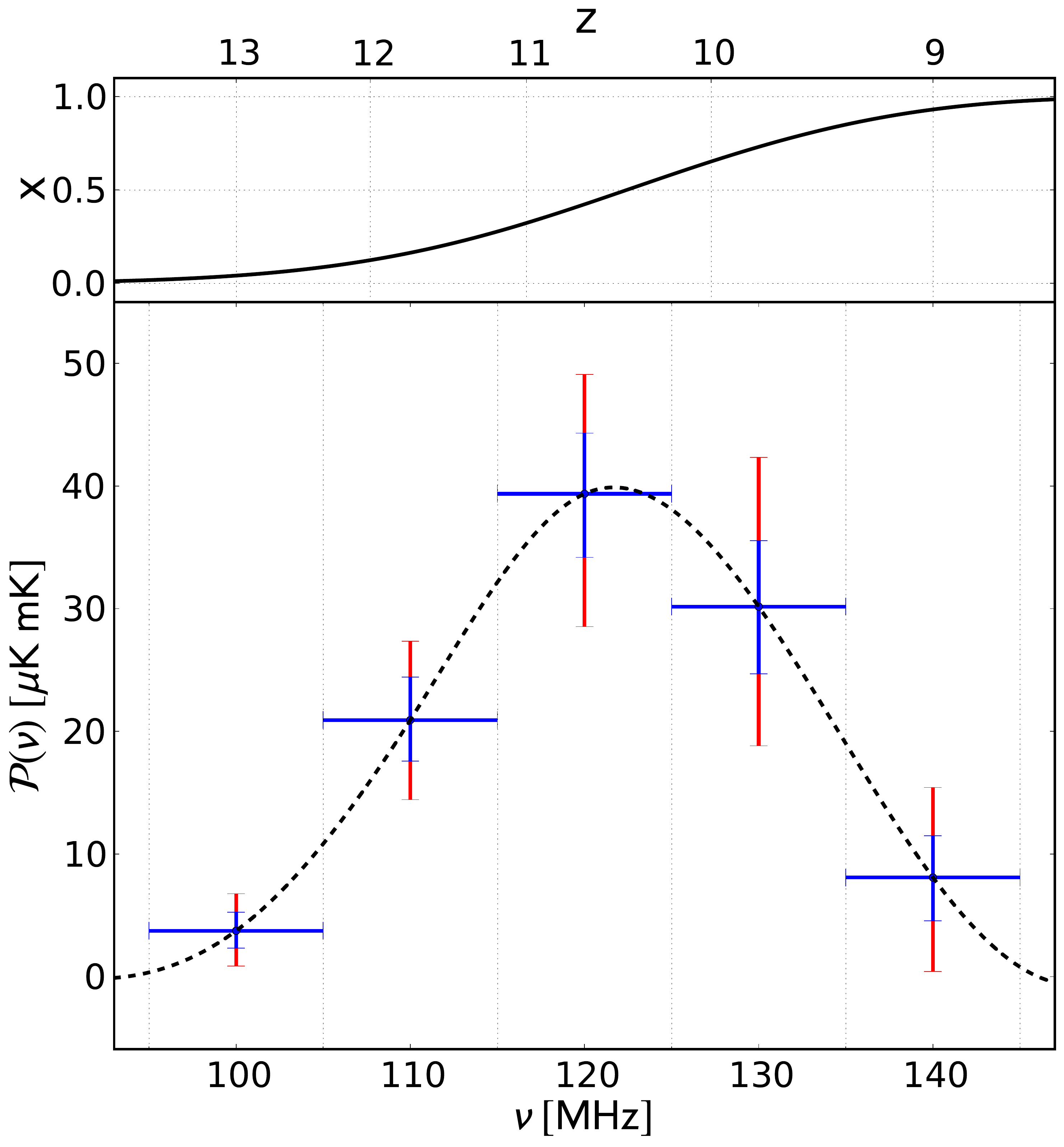}
\vspace{-0.3cm}
\end{center}
\caption{The integrated power up to $\ell=500$, $\sigma^2_{500}$, for 21--cm maps binned in frequency with a bandwidth of $\Delta\nu=20$~MHz.}
\label{fig:sigma_v_nu}
\end{figure}
}

\def\FnOne{%
\footnote{WMAP 2013: ($\Omega_m,\Omega_b,h,n_s,\sigma_8$) = (0.286, 0.0463, 0.69, 0.96, 0.82)
\hfill Planck 2015: ($\Omega_m,\Omega_b,h,n_s,\sigma_8$) = (0.31, 0.0486, 0.68, 0.967, 0.816)} 
}

\def\FnTwo{
\footnote{%
This is a slightly different form from equations (2.9c) and (2.15b) in \citetads{1987ApJ...322..597V}, in which the boundary terms were retained. See the appendix for more details.}
}

\def\FnThree{%
\footnote{The same parametrization as used in CAMB ({\scriptsize\url{http://camb.info}}).}%
}

\def\FnFour{%
\footnote{The boundary terms, evaluated at the lower and upper bounds of the line of sight integral, are omitted, since they generally negligible along the full line of sight. This is not necessarily the case when dealing with simulated maps that only include a range of redshifts. In that case, it can be important to take into account ``surface fluctuations" at the boundaries of the projected region along the line of sight.}%
}

\title{The Kinetic Sunyaev-Zel'dovich Effect from Reionization:\\ Simulated Full Sky Maps at Arcminute Resolution}

\author{Marcelo A. Alvarez}
\affil{Canadian Institute for Theoretical Astrophysics, 60 St George, Toronto ON, Canada, M5S 3H8}

\begin{abstract}
The kinetic Sunyaev-Zel'dovich (kSZ) effect results from Thomson
scattering by coherent flows in the reionized intergalactic medium. We present new results
based on ray-tracing a 10 Gpc scale
simulation at 2-3 Mpc scale resolution to create a full sky kSZ map.
The simulation includes, self-consistently, the effects of reionization on scales
corresponding to multipoles $10\lesssim\ell\lesssim{5000}$. We
separate the kSZ map into Doppler ($\mathbf{v}$), Ostriker-Vishniac
($\delta\mathbf{v}$), patchy ($x\mathbf{v}$), and third-order
($x\delta\mathbf{v}$) components, and compute explicitly all the auto
and cross correlations (e.g., $\langle\mathbf{v}\mathbf{v}\rangle$,
$\langle\delta\mathbf{v}{x}\mathbf{v}\rangle$, etc.) that contribute
to the total power. We find a complex and non-monotonic dependence on
the duration of reionization at $\ell\sim{300}$ and evidence for a
non-negligible (10-30 per cent) contribution from connected four point
ionization-velocity correlations,
$\langle{x}\mathbf{v}x\mathbf{v}\rangle_c$, that are usually neglected
in analytical models. We also investigate the Doppler-large scale
structure (LSS) correlation, focusing on two different probes: (1)
cross power spectra with linearly biased tracers of LSS and (2) cold
spots from infall onto large, rare \ion{H}{2} regions centered on
peaks in the matter distribution at redshifts $z>10$ that are a
generic non-Gaussian feature induced by patchy reionization. Finally,
we use our simulations to show that the reionization history can be
reconstructed at 5-10$\sigma$ significance by correlating full-sky
21-cm maps stacked in bins with $\Delta\nu\!=\!10$ MHz with existing
CMB temperature maps at $\ell\!<\!500$.This raises the prospects of
using more sophisticated velocity reconstruction methods to probe the
distribution of electrons in the IGM by using combined CMB and LSS
measurements well into the epoch of reionization. The resulting kSZ maps 
have been made publicly available at \href{http://www.cita.utoronto.ca/~malvarez/research/ksz-data/}{http://www.cita.utoronto.ca/\~{}malvarez/research/ksz-data/}.
\end{abstract}

\keywords{Cosmic Microwave Background --- Cosmology: Theory --- Large-Scale Structure of Universe
   --- Methods: Numerical}

\section{Introduction}

Secondary anisotropies of the cosmic microwave background (CMB) are an excellent probe of reionization, mainly because scattering of photons by free electrons during reionization affects the observed temperature and polarization in a predictable fashion that is sensitive to the spatial structure of reionization and its correlation with the underlying potential fluctuations. For reviews of the main effects on secondary anisotropies associated with reionization, in the larger context of the effects associated with large scale matter fluctuations, dark matter and gas in halos and filaments, and the relation between cosmic expansion and growth of structure, see \citetads{2002ARA&A..40..171H} and \citetads{2008RPPh...71f6902A}.

Of particular focus in the present work is the kinetic Sunyaev-Zel'dovich (kSZ) effect, which refers to blackbody temperature fluctuations induced by the Doppler shift of CMB photons scattering off of electrons in coherent bulk flows. Although it has long been recognized as one of the most promising probes of the IGM during and after reionization, \citepads[e.g.,]{1978IAUS...79..393S,1984ApJ...282..374K,1986ApJ...306L..51O} it has begun to be used only recently to provide constraints on reionization through the analysis of the angular power spectrum of the CMB temperature at $\ell\approx\!{3000}$ \citepads{2012ApJ...755...70R,2012ApJ...756...65Z,2015ApJ...799..177G}. Significant gains in the accuracy of kSZ power spectrum measurements are expected from new CMB experiments coming online in the near term \citepads[e.g.,]{2014JCAP...08..010C}. A main goal of this work is to refine theoretical predictions for the kSZ effect from patchy reionization and extend existing results to larger scales. Before doing so, we give below a brief historical overview of the development of the field.

Purely blackbody temperature fluctuations arising from the Doppler shift of CMB photons scattered by coherent motions were first discussed by \citetads{1969ApL.....3..189C}, in the context of `whirl' turbulent velocity perturbations on cosmological scales. \citetads{1970Ap&SS...7....3S} also discussed the Doppler effect, but only for velocity fluctuations at recombination; their discussion of anisotropies generated during or after reionization was limited to Comptonization as opposed to coherent Doppler shifting.  The effect from coherent motions of galaxy clusters was first discussed by \citetads{1972CoASP...4..173S}, and has come to be known as the kinetic Sunyaev-Zel'dovich effect. Sunyaev \& Zel'dovich argued that the temperature fluctuation in the Rayleigh Jeans part of the spectrum in the direction of galaxy clusters should be dominated by Comptonization, rather than bulk motion of the cluster itself -- tSZ should dominate over kSZ for individual galaxy clusters, resulting generically in a decrement at low frequencies.

The first appearance in the literature of the secondary temperature anisotropies arising from linear velocity perturbations due to the growth of structure in a diffuse reionized intergalactic medium was in \citeads{1977SvAL....3..268S}, where it was estimated that secondary temperature anisotropies from the Doppler effect could conceivably exceed those generated at recombination, provided that $\tau\sim 1$ and density perturbations on scales corresponding to galaxy clusters were of order unity at reionization. Obviously this is not the case, and the strong sensitivity to the amplitude of perturbations, density of the universe, and timing of reionization was pointed out by \citetads{1978IAUS...79..393S}, where the now well-known line-of-sight Doppler cancellation at small scales was first sketched out. Later calculations, in particular by \citetads{1984ApJ...282..374K}, \citetads{1986ApJ...306L..51O}, and \citetads{1987ApJ...322..597V}, improved greatly on the accuracy of Sunyaev's initial estimates. 

In a somewhat arbitrary adoption of nomenclature, the Doppler effect has come to describe large-scale anisotropies from purely linear velocity perturbations, while the kSZ effect has come to correspond to all secondary anisotropies that depend on electron density fluctuations, including those due to linear and nonlinear gas density fluctuations, individual clusters and groups of galaxies, and the patchiness of reionization. In this paper, the traditional usage of ``Doppler'' will be retained, but ``kSZ" will be used to refer to any blackbody temperature fluctuation arising from bulk motion integrated along the line of sight, {\em including} the Doppler effect. It is in this sense that we refer to the Doppler effect as responsible for large-angle kSZ anisotropies in the rest of the paper.

The power spectrum of kSZ fluctuations is a sensitive probe of the patchiness and duration of reionization at small to intermediate angular scales, $\ell\gsim 300$, due to the transfer of large scale velocity perturbations to small scales by fluctuations in the ionized fraction on scales $1\lsim\!R/{\rm Mpc}\!\lsim 100$ \citepads[e.g.,]{1998ApJ...508..435G,1998PhRvD..58d3001J,1998PhRvL..81.2004K,2001ApJ...551....3G}.  Development in the subject has been mostly theoretical, with numerical simulations playing an increasingly important role in refining the likely shape and amplitude of the angular power spectrum \citepads{2001A&A...367....1V,2004MNRAS.347.1224Z,2005ApJ...630..643M,2007ApJ...660..933I,2010MNRAS.402.2279J,2011MNRAS.414.3424T,2012JCAP...05..007V,2012MNRAS.422.1403M,2013ApJ...769...93P,2013ApJ...776...83B}.

Preliminary analysis by the SPT collaboration \citepads{2012ApJ...755...70R}  determined a 2-$\sigma$ upper--limit on $\dlt$, where $\dl\equiv\ell^2C_\ell/(2\pi)$, of $2.8~\mu$K$^2$ in the case where the thermal SZ--CIB correlation is assumed to be zero, and 6.7~$\mu$K$^2$ when a tSZ--CIB correlation is allowed. After additional observation and analysis, the SPT constraint was considerably tightened. When the bispectrum of the tSZ is used as an additional constraint, SPT data imply $\dlt = 2.9\pm{1.3}$ $\mu$K$^2$ \citepads{2015ApJ...799..177G}. Several authors have discussed the implications of the $\ell\!\approx\!3000$ kSZ measurements for models of patchy reionization, generally interpreting upper limits on $\dlt$ as upper limits on the duration of reionization \citepads{2012ApJ...756...65Z,2012MNRAS.422.1403M, 2013ApJ...776...83B}, although it has been pointed out that this is only generally true for reionization scenarios 
in which the bulk of ionization is from UV photons produced by a quickly--growing population of galaxies in dark matter halos with masses $\gsim 10^{8-9} M_\odot$ \citepads{2013ApJ...769...93P}. 

The kSZ anisotropies on larger scales are much more challenging to detect, due to line of sight cancelation and the dominance of the primary temperature fluctuations \citepads{1978IAUS...79..393S,1984ApJ...282..374K,1987ApJ...322..597V,1994PhRvD..49..648H}, and are thus usually considered entirely negligible \citepads{1995ApJ...439..503D,1998PhRvL..81.2004K,2001A&A...367....1V,2002ARA&A..40..171H,2002PhRvL..88u1301M}. Nevertheless, the kSZ component could be isolated by reconstructing the velocity field on fairly large scales, by using the fact that density and velocity are correlated. Tidal reconstruction in general is a method that is promising due to its quadratic dependence on the underlying field to be estimated, making it less susceptible to systematic effects such as foreground contamination than linear cross-correlations \citepads[e.g.,]{2012arXiv1202.5804P} . The cross-correlation is still important quantity to characterize, however, due to its straightforward interpretation. The most likely tracer of large scale structure in the EOR to be correlated with CMB is the 21--cm background, as first discussed by \citetads{2004PhRvD..70f3509C}, followed by initial attempts at correlating simulated maps by \citetads{2005MNRAS.360.1063S}. 

On large scales ($\ell\sim 100$), where the patchiness of reionization averages out, \citetads{2006ApJ...647..840A} found a substantial CMB--21cm cross-correlation due to the Doppler effect, sensitive to the H~II region bias and reionization history, and first demonstrated that linear matter and kSZ temperature fluctuations induced by peculiar velocity effects are anti-correlated, such that density enhancements during reionization result in cold spots in CMB secondary anisotropies. They found that such a correlation would be detectable with a futuristic experiment like SKA. \citet{2007MNRAS.381..819G} calculated the Doppler--matter cross-correlation in a more general context, confirming the earlier results of \citepads{2006ApJ...647..840A} on the sign and shape of the correlation. \citetads{2008MNRAS.384..291A} \citetads{2010MNRAS.402.2617T} made similar predictions but were pessimistic about its dectecability, due to cosmic variance. However, these studies used simplified analytical expressions for the ionization -- density correlation to estimate the CMB--21cm power spectrum at a given frequency, without stacking frequency maps to increase the strength of the correlation. 

More accurate theoretical predictions for the kSZ fluctuations on intermediate to large angular scales require realistic realizations in large volumes, in order to capture large scale fluctuations in both the ionization field and velocity, which are generally correlated. \citetads{2010MNRAS.402.2279J} carried out simplified radiative transfer simulations in boxes of size 100/$h$~Mpc on a side, corresponding roughly to an angular scale of about a degree. They found that on intermediate scales of $\ell{\sim}10^3$  the correlation will be swamped by the primary CMB fluctuations, while at smaller scales the signal is too small to be detected. They were unable to probe to larger scales, where the cross-correlation is expected to be significantly enhanced, because of the limited simulation volume. The need for more realistic calculations of the kSZ effect and its correlation with the redshifted 21--cm background during reionization served as one of the original motivations for the present work.

As an aside, it is worth mentioning spectral distortions arising from Comptonization of CMB photons in the context of reionization.  Spectral distortions from hot ionized gas are referred to historically as the thermal Sunyaev-Zel'dovich effect \citepads[tSZ;]{1969Ap&SS...4..301Z,1970CoASP...2...66S} and  are most readily detected in the direction of galaxy clusters \citepads{1972CoASP...4..173S}. However, the energy release associated with feedback from early structure formation was in fact the initial motivation for suggesting that deviations from the Planck spectrum in the CMB could result from Compton scattering off of hot electrons at $z<1000$ \citepads{1966ApJ...145..560W}.  Bulk motions along the line of sight, as discussed by \citetads{1972JETP...35..643Z}, produce a nearly identical $y$--type distortion to that produced by Comptonization. The crucial difference being that, in the latter case, the effect depends on the optical depth weighted line-of-sight velocity dispersion rather than electron temperature \citepads{1994PhRvD..49..648H,2004A&A...424..389C}. While such $y$--type spectral distortions could in principle be detected with advanced CMB experiments \citepads[e.g.,]{2015arXiv150701583H}, the present work is limited to blackbody temperature fluctuations produced by coherent motions. 

The outline of this paper is as follows. In \S\ref{sec:ksz} we give a brief review of the basics of the kSZ effect, followed by simplified expressions for the large-scale ($\ell\lsim\!200$) CMB temperature fluctuations expected in currently--favored reionization scenarios, ending with some new estimates of the signature of individual \ion{H}{2} regions in the CMB. In \S\ref{sec:maps} we describe our method of creating kSZ and 21--cm maps from large--scale simulations of patchy reionization, compare to the low--$\ell$ analytical power spectrum derived in \S\ref{sec:ksz}, and perform a novel decomposition of the patchy component into its four constituent terms at both the map and power spectrum level. A discussion of the detectability of the large-angle kSZ--21cm cross-correlation, using the simulated maps, is given in \S\ref{sec:detectability}. The main results are summarized in \S\ref{sec:summary}. 

Velocities are expressed in units of the speed of light, $c=1$, and the explicit relationship between redshift and comoving distance is suppressed, so that $z(\chi)\rightarrow z$ and $\chi(z)\rightarrow \chi$. The approximation $e^{-\tau}\approx\!1$ has been made throughout. Cosmological parameters are based on two flat, $\Lambda{CDM}$ cosmological parameter sets, consistent with results from \citetads[][``WMAP 2013"]{2013ApJS..208...19H} and 
\citetads[][``Planck 2015"]{2015arXiv150201589P}.\FnOne 

\section{Large Scale kSZ from Reionization}
\label{sec:ksz}

In this section, the expressions that determine the kSZ fluctuation along a given line of sight, as well as it's power spectrum, are presented first for an arbitrary electron momentum field separated into curl and divergence--free components. Analytical expressions are obtained using linear theory, and it is shown how the fluctuations created while the ionized fraction is evolving rapidly, as expected during reionization, is qualitatively different from that generated at later times, when the universe is nearly fully--ionized. Finally, an estimate of the size and amplitude of temperature fluctuations created by rare, isolated H~II regions along the line of sight is given.

\subsection{Basic Expressions}
The optical depth to Thomson scattering along a direction $\vgh$ in the sky is given by
\beqa
\tau(\vgh)={\ints}d{\chi}g(z)[1+\delta_e(\chi\vgh)],
\label{eq:tau}
\eeqa
where $\delta_e$ is the electron density contrast, $\delta_e=\delta+\delta_x+\delta_x\delta$,
$\delta$ is the gas density contrast, and the ionization contrast is defined to be $\delta_x= x_e/x-1$, where $x_e{=}n_e/n_{{\rm e},0}$ is the ionized fraction and $x(z)$ is the {\em mean}, volume averaged, ionized fraction, and  $n_{{\rm e},0}=[1-(4-N_{\rm He})Y/4]\Omega_b\rho_{\rm crib}/m_p$, with a helium mass fraction of $Y=0.24$. The number of helium ionizations per hydrogen ionization is set to $N_{\rm He}=1$ and 2 at for $z>3$ and $z<3$, respectively, so that helium is singly ionized along with hydrogen and Helium II reionization occurs instantaneously at $z=3$. The visibility function is defined to be
\beqa
g(z)= \frac{\partial\taua}{{\partial}\chi}= {\sigT}n_{{\rm e},0}x(z)(1+z)^2\equiv g_0(z)x(z),
\eeqa
where $\taua$ is the mean optical depth over the sky. 

The kSZ temperature fluctuation in a direction $\nhat$ is 
\beqa
\frac{\dt}{T_{\rm cmb}}
&=&\ints d\tau~\vel(\chi\nhat)\cdot\nhat = \ints{d}\chi~g(z)~\q(\chi\nhat)\cdot\nhat,
\label{eq:ksz}
\eeqa
where the specific momentum $\q=\left(1+\delta_e\right)\vel=\left(1+\delta\right)\left(1+\delta_x\right)\vel$. This momentum can be separated into transverse and longitudinal components, according to
whether its Fourier transform is perpendicular or parallel to wavenumber, respectively, $\qk_{\perp}=\qk - (\qk\cdot\khat)\khat$ and $\qk_{\parallel}=(\qk\cdot\khat)\khat$.  

The expression for the angular power spectrum is\FnTwo
\beqa
\cl\!=\!\cl^\parallel+\cl^\perp\!=\!
{\ints}d\chi{\ints}d\chi'\left[F_\l^{\perp}(\chi,\chi')+2F_\l^{\parallel}(\chi,\chi')\right],
\label{eq:cl}
\eeqa
where
\beqa
F_\l^i(\chi,\chi')\!&=&\frac{1}{\pi}\!\ints dk~P_{f_i}(k,\chi,\chi')\jl(k\chi)\jl(k\chi'),
\eeqa
$(2\pi)^3P_f(k,\chi,\chi')\delta(\k-\k')=
\left\langle f(\k,\chi) f^*(\k',\chi') \right\rangle$,
and
\beqa
f_\perp(\k,\chi)&\equiv&g(\chi)\qprp(\k,\chi)\\
f_\parallel(\k,\chi)&\equiv&\partial\left[g(\chi)\qpar(\k,\chi)\right]/\partial\chi.
\eeqa 
Note that $\cl^{\parallel}$ is an integral over the power spectrum of the {\em derivative} of $g(z)\q_\parallel$ with respect to comoving distance. It falls off strongly towards small scales  \citepads{1995ApJ...439..503D,1998PhRvL..81.2004K,2001A&A...367....1V,2002ARA&A..40..171H,2002PhRvL..88u1301M} because the visibility function changes slowly along the line of sight, in comparison to the wavelength of the velocity perturbation, leading to nearly complete cancellation \citepads{1978IAUS...79..393S,1984ApJ...282..374K,1987ApJ...322..597V,1994PhRvD..49..648H}. 

\subsection{Angular Power Spectrum of Doppler Effect}
Approximating the velocity using linear perturbation theory, $\vel(\k)=-iaH\!f\delta(\k)\k/k^2$, where $f(z)\!\equiv\!{d}\ln{D}\!/\!d\ln{a}$, and only keeping the leading order term in $q_\parallel(\k)$, so that $f_\parallel(\k)\rightarrow\partial[g(z)\vel(\k)]/\partial\chi$, we obtain
\beqa
\nonumber
F^\parallel_\l\approx\!F^{\parallel{(1)}}_\l&\equiv&\frac{1}{\pi}\frac{\partial{u}}{\partial{\chi}}\frac{\partial{u'}}{\partial{\chi'}}
{\ints}\frac{dk}{k^2}~P(k)\!\jl(k\chi)\!\jl(k\chi')\\
&=&U(\chi,\chi')W_\l(\chi,\chi'),
\label{eq:f1}
\eeqa
where $(2\pi)^3P(k)$ $\delta^D(\k-\k')=$ $\langle \delta(\k)\delta^*(\k')\rangle$,
\beqa
U(\chi,\chi')&\equiv& \frac{\partial{u}}{\partial{\chi}}\frac{\partial{u'}}{\partial{\chi'}}\\
W_\l(\chi,\chi')&\equiv& \frac{1}{\pi}{\ints}\frac{dk}{k^2}~P(k)\!\jl(k\chi)\!\jl(k\chi'),
\eeqa
and
\beqa
\nonumber
u(z)&\equiv& g(z)\dot{D}(z)/D(z)/(1+z)\equiv u_0(z)x(z).
\label{eq:u}
\eeqa

Shown in Figure \ref{fig:dcldzdz} is the contribution per redshift interval to the total signal, 
\beq
\frac{\partial^2{\l^2\cl}}{2\pi\partial{z}\partial{z'}} = \frac{1}{\pi^2}\frac{\partial{u}}{\partial{z}}\frac{\partial{u'}}{\partial{z'}}W_\l(z,z').
\label{eq:dcl}
\eeq
Two cases are shown: uniform ionization (i.e. no recombination, $x=1$; upper--left in each panel), and an analytical reionization history\FnThree given by 
\beq
x(z)=\frac{1}{2}\left[
1+\tanh\left(
\frac{y(z_r)-y}{\Delta{y}}
\right)
\right],
\label{eq:tanh}
\eeq
where $y(z)=(1+z)^{3/2}$ and $\Delta{y}=3/2(1+z_r)^{1/2}\Delta{z}$, and we have set $z_r=10$ and $\Delta{z}=0.5$ (lower--right in each panel).  

There is a stark difference between a reionizing universe, in which the ionization fraction goes from $\sim 0$ to $\sim 1$ over a relatively short interval of redshift, and one in which the ionization fraction is constant, $x=1$. The realistic scenario exhibits a strongly peaked contribution produced at the `surface of re-scattering' at $z\sim{z}_r=10$ (lower right half of each panel). This is due to incomplete cancelation of the line of sight velocity modes, as the ionization fraction evolves significantly even across the velocity perturbation itself \citepads{2006ApJ...647..840A}. The `no-reionization' scenario, however, shows no such peak, and the contribution is much more smoothly distributed with redshift, and confined to relatively larger scales. 

\FigureOne 

For the case of instantaneous reionization, in which the ionized fraction goes from zero to one instantaneously at some redshift $z_*$,
\beqa
\nonumber
U=U_0 + \delta(\!\chi\!-\!\chi_*\!)\delta(\!\chi'\!-\!\chi_*\!)u_0u_0'\!-\!2\delta(\!\chi-\!\chi_*\!)u_0\frac{\partial\!u_0'}{\partial\!\chi'},
\label{eq:u}
\eeqa
where $U_0\equiv(\partial{u_0}/\partial{\chi})(\partial{u'_0}/\partial{\chi'})$.
Combining equations (\ref{eq:cl}), (\ref{eq:f1}) and (\ref{eq:u}), the power spectrum from linear longitudinal velocity fluctuations induced by instantaneous reionization is $C_\l^\parallel=C_\l^{\rm D}+C_\l^{\rm R}-2C_\l^{\rm RD}$, where
\beqa
\nonumber
C_\l^{\rm D}&\equiv&\!{\ints}d\chi\!{\ints}\!d\chi'U_0(\chi\!,\!\chi')W_\l(\chi\!,\!\chi')\\
\nonumber
C_\l^{\rm RD}&\equiv&u_0(z_*){\ints}d\chi\frac{\partial\!u_0}{\partial\!\chi}W_\l(\chi\!,\!\chi_*)\\
C_\l^{\rm R}&\equiv&\!u_0^2(z_*)W_\l^*
\eeqa
and
\beqa
W_\l^*\!=\!\frac{2}{\pi}{\ints}\frac{dk}{k^2}~P(k)\!\jl^2(k\chi_*).
\eeqa
The term $\cl^R$ is corresponds to velocity fluctuations projected onto the re-scattering surface  at $z_*$, while $\cl^D$ corresponds to fluctuations produced from growing velocity fluctuations, i.e. when $x$ is constant and density peaks correspond to hot spots. Shown in Figure \ref{fig:instantaneous} is the angular power spectrum for instantaneous reionization at $z_*=10$, for which $\tau_{\rm es}\simeq 0.09$. A broad peak is evident at $\ell\sim 20-30$, with the re-scattering surface term, $\cl^R$, dominating over the other terms for $\ell\gsim{10}$.

\FigureTwo 

Shown in Figure \ref{fig:cls_anl} are the angular power spectra obtained for $z_r=10$ and $\Delta{z}=0.1, 1$, and 2. All three reionization histories lead to a broad peak at $\ell\sim20-30$, with amplitudes that depend strongly on the duration of reionization. There is a small variation due to changes in the background cosmology. The WMAP 2013 curves are slightly higher for $\ell\lsim\!200$, due at least in part to the slightly redder tilt, $n_s=0.96$ versus $n_s=0.967$ in Planck 2015. The strength of the fluctuations is also linearly dependent on $\Omega_mh^2$ and $\sigma_8^2$, which reflect the strength of the velocity at fixed perturbation amplitude, and the amplitude of perturbations, respectively.  

\FigureThree 
The latest results from the Planck Collaboration indicate a value of $\tau=0.066\pm\!0.013$ at 1$\sigma$ for the ``Planck TT+lowP+lensing+BAO" data combination, with a 2$\sigma$ upper-limit of $\tau=0.092$. When the Planck temperature and polarization data are considered without lensing (``Planck TT+lowP") the constraint changes to $\tau=0.078\pm\!0.019$, implying a 2$\sigma$ upper-limit of $\tau=0.116$. The strong dependence of the large--scale signal on optical depth is shown in Figure \ref{fig:clpk}, where the peak amplitude of the angular power spectrum, $\l_{\rm pk}^2\cl^{\rm pk}/(2\pi)$, is plotted versus $\tau$ for the two fiducial cosmologies calculated. For currently favoured values of $\tau\sim\!0.07-0.08$, the peak values are $\sim\!15-20 \mu{K}^2$, while for $\tau=0.15$ the peak amplitude is expected to lie in the range $60-70 \mu{K}^2$. Its dependence on $\tau$ is approximated with a nearly quadratic function, 
\beq
\frac{\l_{\rm pk}^2\cl^{\rm pk}}{2\pi}\simeq 30 \mu{K}^2 \left(\frac{\tau}{0.1}\right)^{1.9},
\label{fig:fit}
\eeq 
shown as the dashed line in Figure \ref{fig:clpk}. This is quite close to the simple scaling relationship given by \citetads{1984ApJ...282..374K}, $\cl\simeq\langle{v}^2\rangle\tau^2$, just slightly shallower. This is consistent with $\langle{v^2}\rangle$ having a weak inverse dependence on $\tau$, since higher $\tau$ corresponds to somewhat higher redshifts, when the linear velocity perturbations are smaller. 

\FigureFour 

\subsection{Doppler--LSS Cross-correlation}
\label{sec:cc}
The kSZ-LSS cross--correlation can be understood qualitatively by 
examining the behavior of the quantity 
\beq
\frac{\partial{u}}{\partial{z}}=\overline{x}\frac{\partial{u}_0}{\partial{z}}+u_0\frac{\partial{\overline{x}}}{\partial{z}},
\eeq
since 
\beq
\Delta{T}\propto \left(\frac{\partial{u}}{\partial{z}}\right)\left(\frac{\partial{z}}{\partial{\ln\chi}}\right)\delta
\eeq
for a linear density perturbation with amplitude $\delta$ and size $\chi$.  Since in general $\partial{z}/\partial{\ln\chi}>0$, the sign of $\partial{u}/\partial{z}$ determines whether a  linear density enhancement appears as a hot spot or cold spot in the CMB.  When $\partial{u}/\partial{z}>0$, the far side of a density perturbation contributes more, and peaks in the density field will appear as hot spots with $\Delta{T}>0$. This corresponds to when $\partial{x}/\partial{z}\gsim 0$ as is the case during recombination or after reionization. For sufficiently short ionization histories, however, it is possible for the condition
\beq
u_0\frac{\partial{\overline{x}}}{\partial{z}}<-\overline{x}\frac{\partial{u}_0}{\partial{z}}
\eeq
to be satisfied during reionization. In this case, $\partial{u}/\partial{z}<0$, and density enhancements will appear as cold spots, the implications of which will be further discussed in \S\ref{subsec:single}.

Any arbitrary large--scale linearly biased tracer of the matter density can be written as
\beq
f(\nhat)=\ints dz\frac{d\chi}{dz}w(z)\left[b(z)\delta(\chi\nhat)-\frac{1+z}{H(z)}\frac{\partial{\vel}}{\partial\chi}\cdot\nhat\right],
\eeq
where $b(z)$ is the linear bias and $w(z)$ is the redshift weighting. The second factor accounts for the redshift--space distortion in the linear regime \citepads{1987MNRAS.227....1K}.
The cross-correlation between $f(\nhat)$ and $\Delta{T}(\nhat)/T$ is (see derivation of equation \ref{eq:clc} in the Appendix)
\beqa
\cl^{\rm fT}\!&=&\!\frac{2}{\pi}\!\int\!d\chi{w}(z)\!\int\!d\chi'\!\frac{\partial{u'}}{\partial\chi'}\!\int\!dkP(k)\\
&\times&\left[
b(z)\jl(k\chi)-\frac{f}{k^2}\frac{\partial^2j_l(k\chi)}{\partial\chi^2}
\right]\jl(k\chi').
\eeqa

\FigureFive 

\subsection{Doppler Effect from a Single Perturbation}  
\label{subsec:single}
Consider a uniform density perturbation with amplitude $\delta_*$ at some redshift $z_*$, located at a comoving distance $\chi_*\equiv \chi(z_*)$ from the observer along the direction $\nhat_*$, with comoving radius $R_*$. On scales much smaller than the horizon, we can treat cosmological evolution to first order in conformal time across the perturbation. The observer will see a temperature fluctuation along the line of sight given by
\beqa
\nonumber
\frac{\Delta{T}_*}{T_{\rm cmb}}&=&-{\int_{\chi_*-R_*}^{\chi_*+R_*}}~d\chi{g_0}(z)x(z)\vel(\chi\nhat_*)\cdot\nhat_*\\
&\approx&%
-{\int_{-R_*}^{R_*}}~dR%
\left(g_*+Rg_*'\right)
x(z){\rm v}_{\gamma}(R),
\eeqa
where ${\rm v}_\gamma(R)=\vel\left([\chi_*+R]\nhat_*\right)\cdot\nhat_*$, $g_*\equiv g_0(z_*)$, and $g_*'\equiv dg_0(z_*)/d\chi$. 

To obtain the component of the peculiar velocity along the line of sight, one can assume the perturbation has a small amplitude, $\delta\ll1$. Using linear theory in a matter dominated universe, $\nabla\cdot\vel=-\dot{\delta}/(1+z)=-\delta(\dot{D}/D)/(1+z)$, one obtains
\beq
{\rm v}_\gamma(R) = -\frac{\dot{D}(z)}{3D(z)(1+z)}\delta_*R\approx
-\frac{\delta_*}{3}\left(\mathcal{D}_*+R\mathcal{D}_*'\right),
\eeq
where $\mathcal{D}(z)=\dot{D}/D/(1+z)$, and $\mathcal{D}'=d\mathcal{D}/d\chi$.
The assumption that the perturbation itself is at rest with respect to the CMB is likely to be a good approximation on large scales, and certainly true on average.  Thus, photons scattered into the line of sight on the near side of the perturbation, with $\chi<\chi_*$, experience a redshift, $\vel\cdot\nhat>0$, and the temperature fluctuation is given by
\beqa
\frac{{\Delta}T_*}{T_{\rm cmb}}=\frac{\delta_*}{3}
{\int_{-R_*}^{R_*}}dR\left(g_*+Rg_*'\right)\left(\mathcal{D}_*+R\mathcal{D}_*'\right)x(R)R.
\label{eq:dtp}
\eeqa

The largest possible kSZ effect arising from the perturbation occurs if it is ionized  at $z_*$ on a timescale much shorter than its light-crossing time. Such a condition does not violate causality, because reionization is a locally driven process until percolation. In this case, the ionized fraction changes from zero to one very nearly instantaneously  and the ionized fraction along the line of sight is $x\sim 1$ for $R<0$ and $x=0$ otherwise, so that to leading order in $R_*$, equation (\ref{eq:dtp}) simplifies to 
\beqa
\frac{{\Delta}T_*}{T_{\rm cmb}}\approx%
  \frac{g_*\mathcal{D}_*\delta_*}{3}\int^0_{R_*}RdR
=-\frac{g_*\mathcal{D}_*\delta_*R_*^2}{6},
\eeqa
and the associated temperature fluctuation is
\beqa
{\Delta}T_* \approx\!-10~\mu{\rm K}\!\left(\frac{\delta_*}{10^{-2}}\right)\!\left(\frac{1+z_*}{16}\right)^{5/2}\!\left(\frac{R_*}{200~{\rm Mpc}}\right)^2\!,
\label{eq:dtpr}
\eeqa
where matter domination during reionization has been used. Because rapidly expanding H~II regions are expected to be in highly-biased overdense regions, with $\delta>0$, cold spots are a generic imprint from early, rare H~II regions.

\section{Numerical Approach}
\label{sec:maps}

In this section, it is first explained how full-sky kSZ and 21--cm maps are obtained from a high dynamic range simulation of reionization. Maps of the kSZ effect are calculated with and without the linear density fluctuations and patchy ionization fields included, in order to separate first from second order contributions to the kSZ power spectrum and illustrate visually how large scale velocity anisotropies are partially transferred to smaller scales by density and ionization fluctuations. Finally, auto and frequency dependent cross power spectra are presented for the kSZ and 21--cm maps.

The full map making process, including random field generation, the patchy reionization calculation, and light-cone projection, takes about one hour on 128 nodes (8 cores and $\sim$12 GB available RAM each) with a memory footprint of five floats (density contrast, reionization redshift, and linear velocity) for each of the $4096^3$ resolution elements, or 1.25 TB.

The full-sky kSZ map generated according to the procedure outlined below is shown in Figure \ref{fig:asm}. Such a map is only possible using a simulated volume on the order of ten Gpc across, in order to avoid artefacts from repetition of structure along the line of sight and across the sky and missing long wavelength velocity modes. Fluctuations tens of $\mu{K}$ in amplitude are easily seen on scales of 5-10 degrees for the kSZ map, corresponding to the ``Doppler--peak" at $\ell\sim 20-30$. 

\FigureSix

\subsection{Simulations of Patchy Reionization} 
\label{subsec:simmaps}
The simulation is carried out in a periodic box $8$~Gpc$/h$ on a side and with $4096^3$ resolution elements. Once the background cosmology is fixed (WMAP13 is used, see the introduction for parameter values), there are three additional parameters that enter into the calculation: (1) $M_{\rm min}$ -- the minimum halo mass capable of hosting ionizing sources, (2) $\lambda_{\rm abs}$ -- the mean free path to Lyman-limit absorption systems and (3) $\zeta_{\rm ion}$ -- the number of ionizing photons escaping halos per atom \citepads{2012ApJ...747..126A}. The first two are fixed at $M_{\rm min}=10^9M_\odot$ and $\lambda_{\rm abs}=32$~Mpc$/h$, and the efficiency parameter $\zeta_{\rm ion}$ is varied so as to obtain a given value for the Thomson scattering optical depth $\tau=0.1$, $\zeta_{\rm ion}\simeq 4000$. See \citetads{2012ApJ...747..126A} for details on the model and parameter dependence of the reionization history and morphology.  

\subsection{Light cone projection} 
\label{subsec:simmaps}

The first step in obtaining the simulated maps is to generate a random realization of the three dimensional ionization and density fields during reionization in a representative volume on the ligthcone. Such a realization is obtained with the excursion set reionization algorithm of \citetads{2012ApJ...747..126A}, including a mean background level treatment of the opacity due to self-shielded absorption systems and a linear model for the density and, most crucially, velocity fluctuations. 

\FigureSeven 

The kSZ maps are obtained as follows. First, a random realization of the linear density contrast extrapolated to $z=0$, $\delta_0(\r)$ with the observer located at the origin, $\r=0$, is generated. The excursion set algorithm is then applied to obtain a reionization redshift field, $\delta_0(\r) \rightarrow z_{\rm r}(\r)$. The local density contrast at a distance $\chi$ along a line of sight direction $\gh$, $\delta(\chi\gh)=\delta_0(\chi\gh)D(z)$, and the velocity, $\vel(\chi\gh)$, are obtained using Eulerian linear perturbation theory, and the electron fraction is obtained directly from the reionization redshift field, 
\beq
x_e(\chi\gh)=\Theta[z(\chi\nhat)-z_{\rm r}(\chi\nhat)],
\eeq 
where $\Theta$ is the Heavyside function. Finally, equation (\ref{eq:ksz}) is integrated along the line of sight for each pixel center, using the gas density, velocity, and ionized fraction as determined above.

The mean differential brightness temperature at frequency $\nu$ in a  direction $\nhat$ corresponds to the deviation of the observed intensity, $I_\nu$, from that expected for the CMB,
\beq
\dtb(\nhat)\equiv \frac{c^2I_\nu(\nhat)}{2\nu^2k_B} - T_{\rm cmb} \equiv T^\nu_b(\nhat)-T_{\rm cmb},
\eeq
with the intensity measured in the Rayleigh-Jeans part of the spectrum, it can be replaced by brightness temperature in the equation of transfer along the line of sight,
\beq
T^\nu_b(\nhat) = T_{\rm cmb}e^{-\tau_\nu}+T_s(z,\nhat)(1-e^{-\tau_\nu}),
\eeq
with the spin temperature replacing the source function. We adopt the standard scenario for 21--cm in the epoch of reionization, i.e. $T_s\gg{T}_{\rm cmb}$ and a large scale optical depth that is small, $\tau_\nu\ll{1}$. We ignore the effect of redshift--space distortions, which would only boost the strength of the correlation \citepads{2008MNRAS.384..291A}. The final expression is \citepads[e.g.,]{2006ApJ...647..840A}
\beq
\dtb(\nhat) = T_0(z)[1-x_e(\chi\nhat)][1+\delta(\chi\nhat)],
\eeq
where
\beq
T_0(z)=23\ {\rm mK}
\left(
\frac{\Omega_bh^2}{0.02}
\right)
\left[
\left(
\frac{0.15}{\Omega_mh^2}
\right)
\left(
\frac{1+z}{10}
\right)
\right]^{1/2}.
\eeq
See \citetads{2006PhR...433..181F}  and references therein for additional details on the redshifted 21--cm background from reionization. 

\subsection{Maps} 
Shown in Figure \ref{fig:maps} is a $32\times{32}$ field of view from the full-sk y kSZ maps obtained from the 8$h^{-1}$~Gpc box. Four maps are shown, the total signal and its separation into three components, each according to its contribution to the integrand in equation (\ref{eq:ksz}), which can be written
\beqa
\frac{\dt}{T}  
&=&\ints (1+\delta+\delta_x+\delta\delta_x)\ \vel\cdot\nhat\ d\tau.
\label{eq:ksz2}
\eeqa
The ``Doppler" component is the first term $\propto\vel$, while the ``linear OV" and ``patchy" are the second and third terms, $\propto\vel\delta$ and $\propto\vel\delta_x$, respectively.   The Doppler case was obtained by setting the density and ionization fraction fluctuations to zero within the volume, $\delta=\delta_x=0$. The linear OV and patchy maps were obtained by subtracting maps with $\delta=0$ and $\delta_x=0$ from the total, respectively. The fourth component (not shown), $\vel\delta\delta_x$, was obtained by subtracting the linear OV component from the total.

\FigureEight 

The large angle fluctuations on scales of $\sim 5-10$ degrees, with fluctuation amplitudes reaching as large as $30 \mu{K}$, are due nearly entirely to velocity fluctuations. This is seen clearly by comparing the lower-left panel (``Doppler") with the lower-right panel (``total").  Even in the uniform case, where density and ionization fluctuations are completely neglected, the large-scale fluctuation pattern stays essentially the same. Patchiness is certainly an important contribution to the fluctuations, but only on small scales, where it is the dominant source of kSZ fluctuations from reionization (e.g., compare top-right panel of Figure \ref{fig:maps} with $\delta_x=0$ to the lower-left panel with $\delta=0$).

\newcommand{\tpt}[2]{\langle{#1}\!\cdot\!{#2}\rangle_\ell}
\subsection{Deconstructing the kSZ Angular Power Spectrum} 
In what follows it will be useful to refer to the angular power spectrum in terms of the sum of all possible correlations of the individual components of the map. Since the kSZ temperature is given by the optical-depth weighted integral of four terms: $\vel(1+\delta+\delta_x+\delta\delta_x)$ (equation \ref{eq:ksz2}), the power spectrum will be the sum of ten independent power spectra, represented schematically as:
\beqa
\nonumber
\tpt{\Delta{T}}{\Delta{T}} &=& 
\tpt{\vel}{\vel}+
\tpt{\vel\delta}{\vel\delta}+
\tpt{\vel\delta_x}{\vel\delta_x}+
2\tpt{\vel\delta}{\vel\delta_x}\\
\nonumber
&+&2\tpt{\vel\delta}{\vel\delta\delta_x}+
2\tpt{\vel\delta_x}{\vel\delta\delta_x}+
\tpt{\vel\delta\delta_x}{\vel\delta\delta_x}\\
&+&2\tpt{\vel}{\vel\delta}+
2\tpt{\vel}{\vel\delta_x}+
2\tpt{\vel}{\vel\delta\delta_x}.
\label{eq:terms}
\eeqa
The first two terms are easily identifiable as the Doppler and Ostriker-Vishniac effects, respectively. All of the terms were obtained explicitly by cross-correlating the four maps described in \S3.3.

\FigureNine

Shown in Figure \ref{fig:cls_sim} is the angular power spectrum of simulated kSZ map over the full range of multipoles, $3\!\lsim\!\ell\!\lsim\!3000$. In order to focus on the properties of the patchy reionization signal, we have removed the components of the power spectrum that depend solely on gas (as opposed to electron) momentum fluctuations, $\delta\vel$, as $\tpt{\delta\vel}{\delta\vel}$ and $\tpt{\vel}{\delta\vel}$. In practice this amounts to subtracting out the linear Ostriker-Vishniac term,  
\beq
\l^2\cl^{\rm ov} \equiv \frac{1}{2}{\ints}\frac{d\chi}{\chi^2}{u}^2(z)P^{\rm ov}_{\qprp}(\l/\chi),
\label{eq:lov}
\eeq
where
\beq
\nonumber
P^{\rm ov}_{\qprp}(k)\equiv \ints\frac{d^3\k'}{(2\pi)^3}P(|\k-\k'|)P(k')\frac{k(k-2k'\mu)(1-\mu^2)}{k'^2(k^2+k'^2-2kk'\mu)}
\eeq
and $\mu\equiv\khat\cdot\khat'$. That is to say $\tpt{\vel}{\vel\delta}$ is entirely negligible since $\vel$ and $\delta$ are only calculated to linear order in the simulation, and the non--trivial information content is contained in the non--Gaussian $\delta_x$ component. 

The most noticeable feature is the broad peak in the total angular power spectrum at $\ell\sim\!20-30$, corresponding to the Doppler effect. The solid blue line, obtained by calculating the power spectrum from maps in which $\delta=\delta_x=0$, indicates that pure velocity correlations completely dominate the power spectrum at $\ell\lsim{200}$. The dotted blue line lying nearly on top of the solid one is the solution for the longitudinal term in the power spectrum from equation (\ref{eq:cl}), using the linearized mode--coupling term given in equation (\ref{eq:f1}). The only information used from the simulation in calculating the analytical power spectrum is the reionization history, $x(z)$, which enters into equation (\ref{eq:f1}) through $u(z)=\sigma_Tn_{e,0}(1+z)^2\dot{D}(z)/D(z)/(1+z)x(z)$.  This consistency between the analytical formula and ray--tracing result validates the simulation pipeline and shows that density and ionization fluctuations can be neglected for the kSZ power spectrum at $\ell\lsim{200}$, as expected. The agreement between the simulated and analytical power spectrum at $\ell\lsim{200}$ does not depend on whether the OV component  (equation \ref{eq:lov}) is subtracted or not.

The situation becomes more interesting at $\ell\gsim{300}$, as can be seen in Figure \ref{fig:terms}. For clarity, we have also subtracted out the Doppler term, so that solid black line labeled ``Total" corresponds to 
\beqa
\nonumber
\tpt{\Delta{T}}{\Delta{T}} &=& 
\tpt{\vel\delta_x}{\vel\delta_x}+
2\tpt{\vel\delta}{\vel\delta_x}\\
\nonumber
&+&2\tpt{\vel\delta}{\vel\delta\delta_x}+
2\tpt{\vel\delta_x}{\vel\delta\delta_x}+
\tpt{\vel\delta\delta_x}{\vel\delta\delta_x}\\
&+&2\tpt{\vel}{\vel\delta_x}+
2\tpt{\vel}{\vel\delta\delta_x}.
\label{eq:patchy}
\eeqa
Once again, the correlations in which one of the components contains only the projected velocity, $\tpt{\vel}{\vel\delta_x}$ and $\tpt{\vel}{\vel\delta\delta_x}$ are entirely negligible and are everywhere $<0.01\ \mu{\rm K}^2$, consistent with numerical noise. Somewhat surprisingly, the sixth--order term, $\tpt{\vel\delta\delta_x}{\vel\delta\delta_x}$ contributes at about the 5 per cent level at $\ell\lsim{3000}$, while $\tpt{\vel\delta}{\vel\delta\delta_x}$ is entirely negligible. By far the dominant term is the patchy--patchy correlation, $\tpt{\vel\delta_x}{\vel\delta_x}$, accounting for about 65 per cent of the signal at $\ell\!\sim{3000}$. The OV--patchy term, $\tpt{\vel\delta}{\vel\delta_x}$, has a very similar shape to the purely patchy one, but with an amplitude roughly 5 times smaller, perhaps indicating a noisy correlation of ionization fraction with density that persists to small scales, although more work will be necessary to fully understand the patchy--OV term. Finally, $\tpt{\vel\delta_x}{\vel\delta\delta_x}$, while comparable in amplitude to the pathcy--OV term at $\ell\!\sim{3000}$ becomes negative at $\ell\!\lsim\!{900}$, corresponding to a projected comoving radius of $\sim 15-20$~Mpc, quite close to the typical \ion{H}{2}--region scale at the mid--point of reionization \citepads[e.g.,]{2012ApJ...747..126A}.

Also shown in Figure (\ref{fig:terms}) is an approximation for $\tpt{\vel\delta_x}{\vel\delta_x}$ which neglects the connected, or irreducible, fourth moment of the correlation (see equation \ref{eq:xvxva} in the Appendix):
\beq
\l^2\cl^{\langle{x\vel{x}\vel}\rangle} \approx \frac{1}{2}{\ints}\frac{d\chi}{\chi^2}{u}^2(z)P^{\langle{x}\vel\rangle\langle{x}\vel\rangle}_{\qprp}(k)(\l/\chi),
\label{eq:xvxv}
\eeq
where
\beqa
P^{\langle{x}\vel\rangle\langle{x}\vel\rangle}_{\qprp}(k)&\equiv& \ints\frac{d^3\k'}{(2\pi)^3}(1-\mu^2)
\left[
\vphantom{\frac{k'^2P_{x\delta}(|\k-\k'|)P_{x\delta}(k')}{k'^2(k^2+k'^2-2kk'\mu)}}
P(k')P_{xx}(|\k-\k'|)\right.\\
&-&\left.\frac{k'^2P_{x\delta}(|\k-\k'|)P_{x\delta}(k')}{k^2+k'^2-2kk'\mu}
\right].
\eeqa
This term was obtained by tabulating $P_{xx}$ and $P_{x\delta}$ from the simulation for all redshifts such that $0.01\!<\!x(z)\!<\!0.99$ and integrating over the same range of redshift, since there should be no contribution to this term after reionization is complete. The contribution of the connected fourth moment is given by \citepads{2002PhRvL..88u1301M,2015arXiv150605177P} 
\beqa
P^{\langle{x}\vel{x}\vel\rangle_c}_{\qprp}&=&\ints\frac{d^3\k'}{(2\pi)^3}\ints\frac{d^3\k''}{(2\pi)^3}
\sqrt{1-\mu^2}\sqrt{1-\mu'}\\
&\times&\frac{\cos(\phi'-\phi'')}{k'k''}P_{xx\delta\delta}(\k-\k',-\k-\k'',\k',\k'').
\eeqa
In order to contribute significantly, $P_{xx\delta\delta}$ would need to have some dependence on $\phi'-\phi''$ at fixed $\mu'$ and $\mu''$. \citetads{2002PhRvL..88u1301M} pointed out that nonlinearities arising from non-overlapping, static spherical density enhancements -- i.e. the halo model -- exhibit no such dependence on $\phi'-\phi''$, since their internal structure has no explicit correlation with the velocity field. Recently, \citetads{2015arXiv150605177P} pointed out that this is not generally the case, finding a contribution on the order of tens of per cent  lower redshifts, for the late--time kSZ effect, by using perturbation theory and $N-body$ simulations. It seems plausible that even larger departures, such as those seen in Figure \ref{fig:terms}, could occur for such fluctuations induced by patchy reionization, but more detailed analysis will be required before any definitive statements can be made.

\section{CMB--21cm Cross-correlation}
\label{sec:detectability}

In this section we describe the large--scale CMB--21cm cross-correlation, based upon full-sky kSZ and 21--cm maps. We then give some preliminary estimates for its detectability under ideal observing conditions in which instrumental noise and foregrounds can be neglected and the observations are carried out over the full sky. Such observations will be considered to be cosmic variance limited.

\subsection{Cross-correlation Power Spectrum} 
An ideal observation of CMB temperature is the sum of uncorrelated primary and kSZ fluctuations, $\Delta{T}=\Delta{T}_{\rm p}+\Delta{T}_{\rm k}$, with spherical harmonic coefficients $a^T_{\lm}=a^{\rm pp}_{\lm}+a^{\rm kk}_{\lm}$, for which the temperature power spectrum will have a mean value of $\cl^{TT}= \cl^{\rm pp}+\cl^{\rm kk}$ 
and variance $V_\ell^{TT} = 2 \bigl(\cl^{\rm TT}
 \bigr)^2/(\nm)$.

\FigureTen

An ideal observation of 21-cm differential brightness temperature at frequency $\nu$ is $\delta{T}_b^\nu$, with coefficients $a_{\lm}^\nu$ and power spectrum $\cl^\nu\nu$ -- i.e. the 21--cm observation is assumed to contain only the signal from HI, without any noise. The observed cross-correlation will have a mean of $\cl^{T\nu}=\cl^{{\rm k}\nu}$ and a variance of 
\beq 
V_\ell^{T\nu} = \cl^{TT}\cl^{\nu\nu}+(\cl^{T\nu})^2/(\nm)\approx\cl^{\rm TT}\cl^{\nu\nu}(\nm).
\eeq
Defining the auto and cross correlation coefficients as
\beq
\bigl(r_\ell^{kk}\bigr)^2\equiv \bigl(\cl^{kT}\bigr)^2/\ [\cl^{kk}\cl^{TT}] = 
\cl^{kk}/\cl^{TT}
\label{eq:rkk}
\eeq
and
\beq
\bigl(r_\ell^{k\nu}\bigr)^2\equiv \bigl(\cl^{k\nu}\bigr)^2/\ [\cl^{TT}\cl^{\nu\nu}], 
\label{eq:rknu}
\eeq
one obtains a signal to noise of 
\beq
\sn_{kk} = \sum_\ell \bigl(\cl^{kk}\bigr)^2/\ V^{TT}_\ell
= \frac{1}{2}\sum_\ell \nmb\bigl(r_\ell^{kk}\bigr)^2
\label{eq:snrkk}
\eeq
for the kSZ auto-correlation and
\beqa
\nonumber
\sn_{k\nu} &=& \sum_\ell \bigl(\cl^{k\nu}\bigr)^2\ V^{T\nu}_\ell \\
&=& \sum_\ell \nmb
\bigl(r_\ell^{k\nu}\bigr)^2/\ \left[
\bigl(r_\ell^{k\nu}\bigr)^2+1\right]
\label{eq:snrknu}
\eeqa
for the kSZ--21cm cross-correlation.

Shown in Figure \ref{fig:cl21ksz} are the power spectrum, correlation coefficient (equation \ref{eq:rknu}) and cumulative signal to noise ratios of the kSZ--21cm cross-correlation (equation \ref{eq:snrknu}). The two maps show a significant correlation with the same shape as the matter power spectrum, peaking at $\ell\sim 100$, as predicted by \citetads{2006ApJ...647..840A}. The sign of the correlation can be understood in terms of correlated linear velocity and density fluctuations, with an ionized fraction that is strongly biased with respect to the density and smoothly distributed on large scales, hundreds of Mpc across. Although matter and Doppler fluctuations are negatively correlated during reionization (see, e.g., Figure \ref{fig:dudz}), overdense regions contain less neutral hydrogen than underdense regions during reionization because of the bias of ionizing sources, and the overall effect leads to a positive correlation \citepads{2006ApJ...647..840A}. We have verified that the cross-correlation in the case where ionization fraction fluctuations are ignored, $\delta_x=0$, is negative, since the 21--cm map traces in that case traces the matter density (see, e.g., discussion in \S\ref{sec:cc}).

\FigureEleven

\subsection{Dependence on Redshift} 
\label{sec:21vz}
In order to explore the signal-to-noise for a given spectral resolution, a Monte Carlo procedure was adopted. For each trial $i$ of $N$, a random realization of the primary CMB temperature fluctuations, $\Delta{T}_{{\rm p},i}(\nhat)$, was created. The simulated 21--cm and kSZ maps, $\dtb(\nhat)$ and $\Delta{T}_{\rm kSZ}(\nhat)$, were held fixed since the sample variance of the primary CMB fluctuations are independent of the 21--cm signal and dominate the noisiness of the measured cross-correlation. The observed correlation for each trial is given by
\beq
(2\ell+1)C_{\ell,i}^{\rm T\nu} =\sum_{m=-\ell}^{\ell}\!a^{T}_{\lm}a^{{\nu}*}_{\lm,i},
\eeq
where
\beq
a^{\nu}_{\lm} = \ints d\nhat Y_{\lm}(\nhat)\Delta{T}_{21}(\nhat)
\eeq
and 
\beq
a^{T}_{\lm,i} = \ints d\nhat Y_{\lm}(\nhat)[\Delta{T}^{\rm p}_i(\nhat)+\Delta{T}^{\rm kSZ}(\nhat)].
\eeq
If $\Delta{T}_{\rm kSZ}$ and $\Delta{T}_{21}$ are both Gaussian, then it is in principle possible to calculate analytically the distribution of $C_{\ell}^{T\nu}$. However, generating a sufficient number of maps ($N\sim 10^4$) valid for $\ell\lsim 500$, where the correlation is significant, is not a computationally intensive task and includes non-trivial effects of non-Gaussianity due to patchiness in the ionization field and its nonlinear relationship with the underlying Gaussian density fluctuations.

Because the the dependence on frequency, $\nu$ (or, equivalently, redshift, $z$) of the correlation, as opposed to it's dependence on multipole, $\ell$, contains the information that is most difficult to ascertain otherwise, namely the shape of the reionization history, it makes sense to integrate over the multipoles of the correlation at fixed $\nu$. For the frequency binning, one should use as large bins as possible, since that will decrease the instrumental noise (which is neglect in this analysis) and increase the signal in the correlation. Too large a bin, however, and information about its redshift-dependence will be lost in the averaging process. 

Figure \ref{fig:sigma_v_nu} shows the redshift dependence of the integrated power of the cross-correlation in frequency bin $i$,
\beq
\mathcal{P}^{T\nu}(\nu_i) = \sum_\ell \int_{\nu_i-\Delta\nu/2}^{\nu_i+\Delta\nu/2}{w}_\ell(\nu)C_\ell^{T\nu}(\nu){d}\nu,
\eeq
where the weight function, $w_\ell(\nu)$ is normalized to unity.
It can be chosen so as to optimize the overall signal to noise of the cross-correlation \citepads[e.g.,]{2000ApJ...540..605P}. Because the signal depends on a relatively undetermined reionization history, however, it is in practice not possible to know beforehand which weighting function to use, and so $w_\ell(\nu_i)=1/\Delta\nu$ is used, with $\Delta\nu$ the same for all frequency bins.

\section{Summary}
\label{sec:summary}

The angular power spectrum for the kSZ effect due to reionization was calculated on all scales $\ell\lsim\!3000$, using analytical models and numerical simulations of patchy reionization. The angular power spectrum generally exhibits two main features, a broad peak at $\ell\sim\!20-30$ with an amplitude $\sim\!10-30\mu{K}^2$, and a small-scale plateau at $\ell\gsim 300$, with an amplitude of $\sim\!1-5\mu{K}^2$.  The low--$\ell$ peak is caused by the so-called ``Doppler-effect", arising from longitudinal modes in the velocity field ${\rm v}_{\parallel}=(\vel\cdot\khat)$, while the broad plateau at higher multipoles is a superposition of transverse momentum correlations seeded by patchy ionization and density fluctuations, $x\vel$ and $\delta\vel$, respectively.

These results are in good agreement with previous analyses from small scale simulations of patchy reionization, $\ell\gsim 1000$, with a roughly constant amplitude of order a few $\mu{K}^2$ \citepads{2003ApJ...598..756S, 2005ApJ...630..657Z,2005ApJ...630..643M,2007ApJ...660..933I,2011ApJ...727...94T,2012MNRAS.422.1403M,2013ApJ...776...83B,2013ApJ...769...93P}.  For reionization driven by UV sources located in relatively rare dark matter halos -- the scenario favoured by existing data -- the patchiness of reionization is `seeded' by large scale velocity modes, and  the amount of small scale power is dependent mainly on the duration and redshift of reionization, such that more extended reionization histories lead to larger fluctuations at $\ell\sim 3000$ at fixed $\tau$.  

On large scales, $\ell\lsim 300$, where the Doppler effect is the dominant source of kSZ fluctuations, the situation is reversed -- shorter reionization histories lead to larger fluctuations at fixed $\tau$. Unlike the small-scale fluctuations that are created by inhomogeneities in the electron density that quickly lose coherence and accumulate along the line of sight, the large-scale velocities are coherent on large-scales, such that longer reionization histories lead to more cancelation, instead.

At intermediate scales, $300\lsim\ell\lsim{1000}$ the signal exhibits features that could in principle provide additional constraints on reionization from CMB data alone.  For example, it has been recently suggested that one can distinguish between primary and secondary temperature fluctuations by subtracting the theoretical power spectrum for primary fluctuations from the actual observed temperature fluctuations, using independent cosmological parameter constraints obtained from polarization measurements -- what's left would be the kSZ power spectrum \citepads{2014JCAP...08..010C}.  The analysis presented here indicates that such an approach would be quite sensitive to the duration of reionization, since the low-- and high--$\ell$ peak and plataeu, respectively, have opposite dependences on $\Delta{z}$. 

We performed a novel, map--based separation of the patchy kSZ anisotropies on scales $300\lsim\ell\lsim{3000}$, finding several new results. First, the fluctuations from the patchy component at are dominated by correlations of the form $\tpt{\vel\delta_x}{\vel\delta_x}$, which comprise 60 to 70 per cent of the total power over the full range. Surprisingly, the six-point correlation, $\tpt{\vel\delta\delta_x}{\vel\delta\delta_x}$ contributes as much as 5 per cent at small scales, even for the linear density fluctuations we assumed. The term $\tpt{\vel\delta_x}{\vel\delta\delta_x}$ exhibits an interesting feature at $\ell\sim{900}$, where it changes sign. This scale is quite close to the typical radius of \ion{H}{2} regions at the half-ionized epoch, $15--30$~comoving Mpc. Finally, by comparing the leading order patchy term, $\tpt{\vel\delta_x}{\vel\delta_x}$, to the prediction based on the unconnected part of the four--point function, we find tentative evidence for a negative contribution from the connected part at $\ell\lsim{2000}$ and a positive contribution at $\ell\gsim{2000}$. More detailed analysis is required, however, to test this hypothesis and the dependence on the sources and sinks of reionization. 

Going beyond CMB observations alone, the kSZ effect can provide constraints on reionization via cross-correlation with tracers of the large scale density field, particularly the redshifted 21--cm background. At very high redshifts, it's possible that rare, quickly growing \ion{H}{2} regions centered on density peaks could have created large features in the CMB. More work will be required to see just how likely the types of \ion{H}{2} reigon abundances, sizes, growth rates, and biases that are required are to actually occur. At lower redshifts, we confirm with three-dimensional simulations, the prediction of \citetads{2006ApJ...647..840A} that the cross--correlation between the CMB blackbody tempereature and redshifted 21--cm intensity from redshifts $z\sim{10}$ exhibits a broad {\em positive} peak at $\ell\sim{100}$, and that the 21--cm frequency dependence of the correlation traces the reionization history with a resolution of $\Delta{z}\sim{1}$.  Such a correlation is attractive as well because the 21--cm signal naturally contains redshift information, and the systematics of the GHz and MHz band observations are likely to be of quite different origins and therefore uncorrelated.  The difficulty lies however in the `noise' contributed by the much larger primary CMB fluctuations that are peaking at the same multipoles. Indeed, assuming that correlated foregrounds can be cleaned and a sufficiently sensitive instrument can observe over half of the sky, the cross-correlation can be detected at 5--10~$\sigma$ significance, for reionization occuring at $z\sim{10}$ in our fiducial scenario. This raises the prospect of using more precise large scale structure tracers of the velocity, such as large \ion{H}{2} regions and tidal reconstructions, as a new window into the reionization epoch with CMB temperature measurements. 

\acknowledgments
I would like to thank J.~R.~Bond, N.~Battaglia, E.~Komatsu, P.~D.~Meerburg, H.~Park, U.-L.~Pen, and A.~van Engelen for helpful discussions and encouragement. Research in Canada is supported by NSERC and CIFAR. The reionization simulations were performed on the GPC supercomputer at the SciNet HPC Consortium. SciNet is funded by:~the Canada Foundation for Innovation under the auspices of Compute Canada; the Government of Ontario; Ontario Research Fund -- Research Excellence; and the University of Toronto. 
\newpage
\bibliography{ms_ads}
\bibliographystyle{apj}

\newcommand{\asection}[1]{\vspace{0cm}\section{#1}\vspace{0cm}}
\newcommand{\asubsection}[1]{\vspace{0cm}\subsection{#1}}
\renewcommand\thesection{\Alph{section}}
\renewcommand\thesubsection{\Alph{section}\arabic{subsection}}
\setcounter{section}{0} 
\setlength{\parindent}{0pt}
\parskip = 0.5\baselineskip

\section*{Appendix}\vspace{0.15cm}

\renewcommand{\theequation}{A.\arabic{equation}}
\setcounter{equation}{0}

The kSZ angular auto and cross power spectra are derived first for the case where the electron density is spatially homogeneous, where only the longitudinal component of the specific electron momentum contributes, in Appendix A. Appendix B contains a derivation of the angular power spectrum for the transverse component, relevant to the patchy signal, up to leading order in density and ionization fluctuations.

The specific momentum, $\q=(1+\delta_e)\vel$ can be written, to leading order in $\delta$ and $\delta_x$, as
\beq
\q = \vel(1 + \delta + \delta_x)
\eeq
which, when Fourier transformed, is
\beq
\q(\k)=\vel(\k)+\ints\frac{d^3\k'}{(2\pi^3)}
\left[\delta(\k-\k')+\delta_x(\k-\k')\right]\vel(\k').
\label{eq:qft}
\eeq
The momentum is further decomposed into longitudinal and transverse components using
 $\qk_\perp=\q-(\q\cdot\hat{\k})\hat{\k}$,
 \beqa
\qprpv(\k)&=&\ints\frac{d^3\k'}{(2\pi)^3}\left[\delta(\k-\k')+\delta_x(\k-\k')\right]
\velm(\k')\left[\hat{\k}'-\mu'\hat{\k}\right]
\eeqa
and
\beqa
\qpar(\k)&=&\velm(\k)+\ints\frac{d^3\k'}{(2\pi)^3}\left[\delta(\k-\k')+\delta_x(\k-\k')\right]
\velm(\k')\mu'\hat{\k}
\label{eq:qprp}
\eeqa
where $\mu'\equiv \hat{\k'}\cdot\hat{\k}$ and $\vel(\k)=\velm(\k)\khat$ holds on all scales of interest.

\section{Longitudinal Momentum Fluctuations}

The leading order contribution to the fluctuations arising from $\qpar$ depend only on the linear velocity, $\vel=|\vel|\khat$. This greatly simplifies calculation of the momentum fluctuations that contribute to the Doppler component. On large scales, however, the Limber approximation breaks down and the full integral over pairs of lines of sight must be retained.

Fluctuations on the sky of the kSZ temperature can be expressed in
terms of spherical harmonics $C_\ell=\langle a_{\lm}a_{\lm}^*\rangle$,
where $a_{\lm} = \ints d\nhat\ Y_{\lm}(\nhat)\dt/T$.
The kSZ temperature fluctuation can be written in terms of the Fourier transform of the momentum,
\beqa
\dt=\ints{d}\chi~g(z)\q\cdot\nhat\ = \ints{d}\chi~g(z)
\ints\d3k e^{-i\k\cdot \chi\nhat}
\left(
\q_{\perp}\cdot\nhat+\qpar\nhat\cdot\khat
\right) \equiv \dtprp+\dtpar,
\eeqa
from which it follows that $
\cl=\langle a^{\perp}_{\lm}a^{\perp *}_{\lm}\rangle+
\langle a^{\parallel}_{\lm}a^{\parallel *}_{\lm}\rangle 
\equiv \cl^\perp+\cl^\parallel$.
The parallel component is given by:
\beqa
\frac{\Delta{T}_\parallel}{T}=
\int\!d\chi{g}(z)\ints\d3k \qpar(\k)\nhat\cdot\khat
e^{-i\k{\cdot}\chi\nhat}
=\ints\d3k \frac{1}{k}\int\!\qpar(\k)g(z)\frac{{\partial}e^{-i\k{\cdot}\chi\nhat}}{{\partial}\chi}d\chi.
\label{eq:dtpar}
\eeqa
Integrating by parts\FnFour, one obtains
\beqa
\nonumber
\frac{\Delta{T}_\parallel}{T}=-\ints\d3k\frac{1}{k}
{\ints}d\chi{e}^{-i\k{\cdot}\chi\nhat}f_\parallel(\k),
\eeqa
where $f_\parallel(\k)\equiv\partial\left[\qpar(\k)g(z)\right]/\partial\chi$,
Using one of Rayleigh's identities, 
\beqa
\nonumber
\frac{\Delta{T}_\parallel}{T}\!=-4{\pi}\sum_{\lm}i^lY_{\lm}(\nhat)\!\ints\d3k\frac{Y^*_{\lm}(\khat)}{k}\!
{\ints}d\chi{f}_\parallel(\k){j}_l(k\chi)
\eeqa
and the spherical harmonic coefficients are therefore
\beqa
a_{\lm}^\parallel = 4\pi(-\!i)^\ell\!\int\!d\chi\!\int\!\d3k\frac{f_\parallel(\k)}{k}{j}_l(k\chi)Y_{\lm}^*(\khat).
\label{eq:alm}
\eeqa
The angular power spectrum of fluctuations due to the longitudinal component as
\beq
\cl^\parallel =
\langle
a_{\lm}^\parallel a_{\lm}^{{\parallel}*}
\rangle =(4\pi)^2{\ints}\d3k{\ints}\d3k\frac{
\langle
f_\parallel(\k)f_\parallel^*(\k)
\rangle
}{k^2}=
\frac{2}{\pi}
\left[
{\ints}d\chi{\ints}d\chi'
{\ints}dk~\pfppp(k,\chi,\chi')\jl(k\chi)\jl(k\chi')
\right],
\eeq
where $
(2\pi)^3\pfppp(k,\chi,\chi')\delta(\k-{\bm k'})=
\left\langle f_\parallel(\k,\chi) f_\parallel({\bm k'},\chi')^* \right\rangle$.
This is the $\cl^\parallel$ term in equation (\ref{eq:cl}) and the main result of this section. 

The Limber approximation can be obtained by using the relationship $\ell^2\!\ints{P}(k)\jl(k\chi)\jl(k\chi')dk \approx
\pi\delta(\chi-\chi')P\left(\l/\chi,\chi\right)/2$,
which implies
\beq
\l^2\cl^\parallel = \ints d\chi{P}_{f_\parallel}(k=\l/\chi,z).
\label{eq:cpar}
\eeq
This is only a good approximation when $\l\gg{H}(z)\chi(z)/\delta{z}$, where $\delta{z}=f_\parallel/(\partial{f_\parallel}/dz)$.

\renewcommand{\theequation}{B.\arabic{equation}}
\setcounter{equation}{0}
\asection{Transverse Momentum Fluctuations}

The Limber approximation is generally valid at multipoles where the transverse momentum fluctuations contribute to kSZ fluctuations. Because the velocity field is longitudinal, however, the momentum fluctuations are second order, and depend on the density and ionization fluctuations and their cross-correlations with each other and the velocity. 

The angular power spectrum of the transverse component is given accurately by Limber's approximation \citepads[e.g.,]{2013ApJ...769...93P},
\beqa
\nonumber
\cl^\perp = \frac{1}{2}\ints \frac{d\chi}{\chi^2}g^2(z)P_{\qprp}(k=\l/\chi,z),
\label{eq:clperp}
\eeqa
where $(2\pi)^3P_{\qprp}(k,z)\delta^D(\k-\k')=
\langle
\q_\perp(\k)\q_\perp(\k')^*
\rangle$. Analytical expressions have been derived previously for the $\qprp$ terms in the angular power spectrum \citepads[e.g.,]{2002PhRvL..88u1301M,2003ApJ...598..756S,2013ApJ...769...93P}. \citetads{2002PhRvL..88u1301M} derived an expression for the power spectrum of $\q_\perp(\k)$ in linear theory (their equation 7), but neglecting patchiness (i.e. $\delta_x=0$ in equation \ref{eq:qprp}). In what follows we derive the expression for the power spectrum with patchy terms included.

The power spectrum can be obtained from equation \citepads[\ref{eq:qft} -- see also][]{2005ApJ...630..643M,2000ApJ...529...12H}
\beqa
\langle
\q_\perp(\k)\q_\perp(\k)^*
\rangle
&=& \int\!\frac{d^3\k'}{(2\pi)^3}\int\!\frac{d^3\k''}{(2\pi)^3}
\left\langle
\left[\delta(\k'_0)+\delta_x(\k'_0)\right]
\velm(\k')
\left[\delta^*(\k''_0)+\delta^*_x(\k''_0)\right]
\velm^*(\k'')(\hat{\k}'-\mu'\hat{\k})\cdot(\hat{\k}''-\mu''\hat{\k})
\right\rangle,
\eeqa
where $\k'_0\equiv\k-\k'$ and $\k''_0\equiv\k-\k''$. In general, a four-point function can be written 
\beq
\langle ABCD\rangle = \langle AB\rangle\langle CD\rangle+\langle AC\rangle\langle BD\rangle+\langle AD\rangle\langle BC\rangle+\langle ABCD\rangle_c,
\eeq
where $\langle ABCD\rangle_c$ is the irreducible or connected fourth moment. It has been pointed out that the connected term is generally sub-dominant with respect to the others at late times, due to sphericalization of density perturbations in the nonlinear regime \citepads{2002PhRvL..88u1301M}, with a contribution on the order of ten percent \citepads{2015arXiv150605177P}. Isotropy implies that terms containing factors with the form $\langle f(\k-\k')g(\k')^*\rangle$ are only non-zero for $\k=0$, so  only terms containing combinations of factors of the form $\langle f(\k-\k')g(\k-\k'')^*\rangle$, $\langle f(\k')g(\k'')^*\rangle$, $\langle f(\k')g(\k-\k'')^*\rangle$, and $\langle f(\k'')g(\k-\k')^*\rangle$ remain. The resulting expression is
\beqa
P_{\qprp}(k) &=& 
\frac{1}{(2\pi)^2}\int\!k'^2\!dk'\!\int\!d\mu(1-\mu^2)
\left\lbrace
P_{vv}(k')
\left[
P_{\delta\delta}(|\k-\k'|)+
P_{xx}(|\k-\k'|)+2P_{x\delta}(|\k-\k'|)
\right]
\right. -
\\
&&
\left.
\frac{k'}{|\k-\k'|}\left[
P_{\delta{v}}(|\k-\k'|)P_{v\delta}(k')+
P_{\delta{v}}(|\k-\k'|)P_{xv}(k')+
P_{xv}(|\k-\k'|)P_{\delta{v}}(k')+
P_{xv}(|\k-\k'|)P_{xv}(k')
\right]
\right\rbrace,
\eeqa
where $|\k-\k'|=(k^2+k'^2-2kk'\mu)^{1/2}$.
The velocity obtained from linear perturbation theory is an excellent approximation and gives $P_{\delta{v}}(k)=\alpha(z)P_{\delta\delta}(k)/k$, $P_{vv}(k)=\alpha^2(z)P_{\delta\delta}(k)/k^2$, and $P_{xv}(k)=\alpha(z)P_{x\delta}(k)$, where $\alpha(z) \equiv \dot{D}(z)/D(z)/(1+z)$. The expression is further simplified to
\beqa
P_{\qprp}(k) &=& 
\frac{\alpha^2}{(2\pi)^2}\int\!dk'\!\int\!d\mu(1-\mu^2)
\left\lbrace
P_{\delta\delta}(k')
\left[
P_{\delta\delta}(|\k-\k'|)+
P_{xx}(|\k-\k'|)+2P_{x\delta}(|\k-\k'|)
\right]
\vphantom{ \int\!\frac{d^3\k'}{(2\pi)^3}}
 \right. -
\\
&&
\left.
\frac{k'^2}{|\k-\k'|^2}\left[
P_{\delta\delta}(|\k-\k'|)P_{\delta\delta}(k')+
P_{\delta\delta}(|\k-\k'|)P_{x\delta}(k')+
P_{x\delta}(|\k-\k'|)P_{\delta\delta}(k')+
P_{x\delta}(|\k-\k'|)P_{x\delta}(k')
\right]
\right\rbrace.
\eeqa

Finally, we group the terms by their dependence on density and ionization fluctuations, 
\beq
P_{\qprp}(k) = \frac{\alpha^2}{(2\pi)^2}\int\!dk'\!\int\!d\mu(1-\mu^2)\left[
I_{\qprp}^{\delta\delta}(k,k',\mu) + I_{\qprp}^{xx}(k,k',\mu) + I_{\qprp}^{x\delta}(k,k',\mu)
\right],
\eeq
where 
\beqa
I_{\qprp}^{\delta\delta}(k,k',\mu) &\equiv& 
P_{\delta\delta}(k')P_{\delta\delta}(|\k-\k'|)\frac{k(k-2k'\mu)}{k^2+k'^2-2kk'\mu}
\eeqa
depends only on matter fluctuations ($\langle\vel\delta\vel\delta\rangle$),
\beqa
I_{\qprp}^{xx}(k,k',\mu) &\equiv&
P_{\delta\delta}(k')
P_{xx}(|\k-\k'|)-
\frac{k'^2
P_{x\delta}(|\k-\k'|)P_{x\delta}(k')
}{k^2+k'^2-2kk'\mu}
\label{eq:xvxva}
\eeqa
depends only on patchiness ($\langle\vel\delta_x\vel\delta_x\rangle$), and
\beqa
I_{\qprp}^{x\delta}(k,k',\mu) &\equiv&
2P_{\delta\delta}(k')P_{x\delta}(|\k-\k'|) -
\frac{k'^2\left[
P_{\delta\delta}(|\k-\k'|)P_{x\delta}(k')+
P_{x\delta}(|\k-\k'|)P_{\delta\delta}(k')
\right]}{k^2+k'^2-2kk'\mu}
\eeqa
depends on the cross terms ($\langle\vel\delta\vel\delta_x$). When the ionization field is assumed to be uniform, i.e. $x = \langle{x}\rangle \implies P_{x\delta}=P_{xx}=0$, the only term remaining is $I^{\delta\delta}_{\qprp}$, corresponding to the ``linear Ostriker-Vishniac effect", i.e. $\langle(\delta{v})(\delta{v})'\rangle$. 

\renewcommand{\theequation}{C.\arabic{equation}}
\setcounter{equation}{0}
\asection{Large--scale Cross--correlation with linearly biased tracers}

Consider a projected map of some biased tracer of large--scale structure,
\beq
f(\nhat)\!=\!\int\!d\chi{w}(z)\!\left[b(z)\delta(\chi\nhat)-\frac{1}{aH}\frac{\partial}{\partial\chi}{\vel}\cdot\nhat\right],
\label{eq:f}
\eeq
where $b(z)$ is the linear bias and $w(z)$ is a redshift weighting that encodes the properties of the particular survey. The second factor in the integral accounts for the redshift space distortion from linear velocity fluctuations at first order in the density fluctuation \citepads{1987MNRAS.227....1K}. We derive in what follows an expression for the angular power spectrum of the cross-correlation with the longitudinal (i.e. Doppler) component of the kSZ signal, 
$\Delta{T}_\parallel(\nhat)/T$, as defined in equation (\ref{eq:dtpar}). The cross--correlation in this limit is given by $\cl^{fT}=\langle\alm^f\alm^{\parallel*}\rangle$. Fourier transforming $\delta$ and $\vel$ in equation (\ref{eq:f}), and using $\vel(\k)=-i\k\delta(\k)\dot{D}(z)/[D(z)k^2(1+z)]=-ifaH\delta(\k)\k/k^2$, where $f\!=\!d\ln{D}/d\ln{a}$, we obtain
\beqa
\nonumber
f(\nhat)\!&=&\!\int\!d\chi{w}(z)\!\int\!\d3k\!\delta(\k)\!\left[
b(z)e^{-i\chi\k\cdot\nhat}
+ \frac{if\k\cdot\nhat}{k^2}\frac{\partial}{\partial\chi}e^{-i\chi\k\cdot\nhat}
\right]=
\!\int\!d\chi{w}(z)\!\ints\d3k \delta(\k)\!\left[
b(z)e^{-i\chi\k\cdot\nhat}
- \frac{f}{k^2}\frac{\partial^2}{\partial\chi^2}e^{-i\chi\k\cdot\nhat}
\right]\\
\!&=&\!-4{\pi}\sum_{\lm}i^lY_{\lm}(\nhat){\ints}d\chi{w}(z)\!\ints\d3k{Y}^*_{\lm}(\khat)\!\delta(\k)
\left[
b(z){j}_l(k\chi) - \frac{f}{k^2}\frac{\partial^2j_l(k\chi)}{\partial\chi^2}
\right].
\eeqa
The spherical harmonic coefficients are obtained in the same way as in equation (\ref{eq:alm}), 
\beq
a_{\lm}^f = 4\pi(-\!i)^\ell\!\int\!d\chi{w}(z)\!\int\!\d3k\delta(\k)\left[
b(z){j}_l(k\chi)-\frac{f}{k^2}\frac{\partial^2j_l(k\chi)}{\partial\chi^2}
\right]Y_{\lm}^*(\khat).
\label{eq:almf}
\eeq
The cross-correlation is obtained by using equation (\ref{eq:almf}) and the first order expression $f_\parallel(\k)\rightarrow\partial[g(z)\vel(\k)]/\partial\chi$ in equation (\ref{eq:alm}):
\beq
\cl^{fT}=\langle\alm^f\alm^{\parallel*}\rangle=\frac{2}{\pi}
{\ints}d\chi{w}(z){\ints}d\chi'\frac{\partial{u}}{\partial\chi'}
{\ints}dk~P(k)\left[
b(z)\jl(k\chi)-\frac{f}{k^2}\frac{\partial^2j_l(k\chi)}{\partial\chi^2}
\right]\jl(k\chi').
\label{eq:clc}
\eeq

\end{document}